\documentclass[12pt,pdftex]{article}
\usepackage[dvips]{graphicx}
\usepackage{amsmath,amssymb,amsthm}
\usepackage{mathbbol,bbm}
\usepackage[final]{showkeys}
\usepackage{bm}
\usepackage{calc}
\usepackage{color}

\newcommand{\chtpi}{{\color[rgb]{0.4,0.2,0.9} \sc Qbook:Ch.1}} 
\newcommand{\chpqtc}{{\color[rgb]{0.4,0.2,0.9} \sc Qbook:Ch.2}} 
\newcommand{\chqfs}{{\color[rgb]{0.4,0.2,0.9} \sc Qbook:Ch.3}}
\newcommand{\chsaqt}{{\color[rgb]{0.4,0.2,0.9} \sc Qbook:Ch.4}}
\newcommand{\chpiqa}{{\color[rgb]{0.4,0.2,0.9} \sc Qbook:Ch.5}}
\newcommand{\chqtnt}{{\color[rgb]{0.4,0.2,0.9} \sc Qbook:Ch.6}}
\newcommand{\chqbat}{{\color[rgb]{0.4,0.2,0.9} \sc Qbook:Ch.7}}

\newcommand{\chqfluct}{{\color[rgb]{0.4,0.2,0.9} \sc Qbook:Ch.9}} 
\newcommand{\chflqwork}{{\color[rgb]{0.4,0.2,0.9} \sc Qbook:Ch.10}}

\newcommand{\chqtraj}{{\color[rgb]{0.4,0.2,0.9} \sc Qbook:Ch.14}}
 
\newcommand{\chdtyp}{{\color[rgb]{0.4,0.2,0.9} \sc Qbook:Ch.16}}
 
\newcommand{\chnmbq}{{\color[rgb]{0.4,0.2,0.9} \sc Qbook:Ch.18}} 
\newcommand{\chpots}{{\color[rgb]{0.4,0.2,0.9} \sc Qbook:Ch.19}}

\newcommand{\chreacc}{{\color[rgb]{0.4,0.2,0.9} \sc Qbook:Ch.22}}
\newcommand{\chheom}{{\color[rgb]{0.4,0.2,0.9} \sc Qbook:Ch.23}} 
\newcommand{\chcool}{{\color[rgb]{0.4,0.2,0.9} \sc Qbook:Ch.24}}
\newcommand{\chrto}{{\color[rgb]{0.4,0.2,0.9} \sc Qbook:Ch.25}}

\newcommand{\chthinf}{{\color[rgb]{0.4,0.2,0.9} \sc Qbook:Ch.33}}

\newcommand{\chsion}{{\color[rgb]{0.4,0.2,0.9} \sc Qbook:Ch.36}}

\begin{document}

\begin{center}
{\LARGE Introduction to Quantum Thermodynamics: History and Prospects }\\[18pt]
Robert Alicki\\[6pt]
Institute of Theoretical Physics and Astrophysics \\
University of Gda\'nsk, Poland \\[6pt]
Ronnie Kosloff\\[6pt]
The Fritz Haber Research Center for Molecular Dynamics\\
The Institute of Chemistry \\
The Hebrew University of Jerusalem\\
Jerusalem 91904, Israel \\[6pt]
\end{center}

\begin{abstract}
Quantum Thermodynamics is a continuous dialogue between two independent theories: Thermodynamics and Quantum Mechanics.
Whenever the two theories have addressed the same phenomena new insight has emerged. 
We follow the dialogue from equilibrium Quantum Thermodynamics and the notion of entropy and entropy inequalities which are the base of the II-law.
Dynamical considerations lead to non-equilibrium thermodynamics of quantum Open Systems. 
The central part played by completely positive maps is discussed leading to the Gorini-Kossakowski-Lindblad-Sudarshan "GKLS" equation.
We address the connection to thermodynamics through the system-bath weak-coupling-limit WCL leading to dynamical versions of the I-law.
The dialogue has developed through the analysis of quantum engines and refrigerators. Reciprocating and continuous engines are discussed.
The autonomous quantum absorption refrigerator is employed to illustrate the III-law.
Finally, we describe some open questions and perspectives.
\end{abstract}

\section{Introduction}

Quantum mechanics was conceived from a consistency argument on the nature of thermal emitted light.
In 1900, Planck, as an act of despair, introduced a fix to 
the frequency distribution law of black body radiation \cite{planck1900}. 
In 1905 Einstein  reanalysed the problem,  based on consistency with thermodynamics, he writes:
{\em In terms of heat theory monochromatic radiation of low density (within the realm of validity of Wien's radiation
formula) behaves as if it consisted of independent energy quanta of the magnitude $h \nu$. }
Einstein's conclusion is a quantised electromagnetic field \cite{einstein05} the dawn of quantum mechanics.
From this point on, quantum mechanics developed independently eventually setting its own set of assumptions \cite{vNeumann}.
Currently, the consistency argument is  used in reverse, deriving the laws of thermodynamics from the established quantum principles.
This approach allows naturally the addition of  dynamical out of equilibrium considerations.

In 1916 Einstein  examined the relation between  stimulated emission and radiation absorption
using thermodynamical equilibrium arguments \cite{einstein1916}. 
This paper addressing the light matter interaction is the prerequisite for the theory of Lasers.
Lasers represent a non equilibrium phenomena where amplified light is generated from a non equilibrium distribution of matter. 
In 1959, during the early development of solid state lasers, Scovil and Schulz-Dubios  realized the
equivalence of a three-level maser with a Carnot heat engine \cite{scovil1959}. This is a seminal paper in contemporary quantum thermodynamics.
They identified the amplified light as work and the kinetic process that establishes the population inversion as heat generated by a hot and cold bath
of different temperatures. The well known thermodynamical viewpoint that an engine can be reversed to a heat pump led Geusic, Scovil and Schulz-Dubios 
to suggest Maser cooling \cite{geusic1959} and  in 1967 Laser cooling in the summarizing paper {\em quantum equivalence of Carnot cycle} \cite{geusic1967quantum}.
These studies preceded the work of Wineland and H\"ansch  which reinvented laser cooling in 1975 \cite{wineland1978,hansch1975} which was not based on
thermodynamical arguments.

Thermodynamics is usually viewed as a theory of large scale macroscopic processes. In view of the trend toward miniaturization, how far down
can thermodynamics be applicable? J von Neumann set the foundation for quantum theory on a probabilistic footing relevant for a single particle.
Thus quantum mechanics enables thermodynamical ideas to be applicable on any scale.

\section{Equilibrium Quantum Thermodynamics}

\subsection{The von Neumann mathematical formalism of quantum statistical physics}

A mathematically precise framework of quantum mechanics for systems of finite number of degrees of freedom has been developed by J. von Neumann
in his book \cite{vNeumann}. He synthesized the contribution of E. Schr\"odinger, W. Heisenberg and P.A.M. Dirac in the language of Hilbert spaces and linear operators acting on them \cite{heisenberg1949physical,schrodinger1926undulatory,dirac1981principles}. 
von Neumann established  the following fundamental structure of \emph{quantum probability} \cite{vNeumann}:\\
i) quantum observables are self-adjoint (hermitian) operators  (denoted by $\hat{A}, \hat{B},..$)  acting on the Hilbert space $\mathcal{H}$,\\
ii) quantum events are the particular \emph{yes-no} observables described by projectors ($\hat{P} = \hat{P}^2 $),\\
iii) quantum probability measures are represented by density matrices, i.e.  positive operators with trace one (denoted by  $\hat{\rho}, \hat{\sigma},..$),\\
iv) probability of the event $\hat{P}$ for the state  $\hat{\rho}$ is given by 
\begin{equation} 
\mathcal{P} = \mathrm{Tr}(\hat{\rho} \hat{P}),
\label{event}
\end{equation}
v) an averaged value of the observable $\hat{A}$ at the state $\hat{\rho}$ is equal to 
\begin{equation} 
\langle \hat{A} \rangle_{\rho} = \mathrm{Tr}(\hat{\rho} \hat{A}).
\label{average}
\end{equation}
The reversible dynamics of a quantum system formulated in terms of density matrices is governed by the von Neumann evolution equation with the generally time-dependent Hamiltonian $\hat{H}(t)$ 
\begin{equation} 
\frac{d}{dt}{\hat{\rho}(t)} = -\frac{i}{\hbar}[\hat{H}(t) , \hat{\rho}(t)] ,
\label{vN_eq}
\end{equation}
with the solution in terms of the unitary propagator $\hat{U}(t, t_0)$
\begin{equation} 
\hat{\rho}(t) = \hat{U}(t, t_0) \hat{\rho}(t_0)\hat{U}(t, t_0) ^{\dagger},\quad  \hat{U}(t, t_0) = \mathbb{T}\exp\Bigl\{-\frac{i}{\hbar}\int_{t_0}^t \hat{H}(t') dt'\Bigr\}.
\label{vN_solution}
\end{equation}
where $\mathbb{T}$ is the time ordering operator.
von Neumann introduced also the notion of entropy of the density matrix, called now von Neumann entropy and defined by the expression
\begin{equation} 
S_{vn}(\hat{\rho}) = - k_B \mathrm{Tr}(\hat{\rho} \ln\hat{\rho}) =  - k_B\sum_{j} \lambda_j\ln \lambda_j
\label{entropy}
\end{equation}
where $\hat{\rho}= \sum_{j} \lambda_j |j\rangle\langle j|$ is a spectral decomposition of the density matrix. Notice, that this entropy is well-defined and non-negative (albeit can be infinite) in contrast to the generally ill-defined Boltzmann entropy for classical probability distributions on phase-spaces. The von Neumann entropy is an invariant of the state $\hat \rho$ 
and is the lower bound for $S_A(\hat \rho) \ge S_{vn} (\hat \rho)$ where $S_A = - k_B\sum_{j} p_j \ln p_j$ 
is the Shannon entropy defined by the 
probability distribution obtained by a complete measurement of the operator $ \hat A$.
\par
The quantum counterpart of the canonical (Gibbs) ensemble, corresponding to the thermodynamic equilibrium state at the temperature $T$, for the system with the Hamiltonian $\hat{H}$, is described by the density matrix of the form 
\begin{equation} 
\hat{\rho}_{\beta} =  \frac{1}{Z} e^{-\beta \hat{H}} , \quad \beta = \frac{1}{k_B T} ,\quad Z =  \mathrm{Tr} e^{-\beta \hat{H}} .
\label{gibbs}
\end{equation}
The Gibbs state maximizes entropy under the condition of a fixed mean energy (internal energy in thermodynamic language) 
$E =\mathrm{Tr}(\hat{\rho} \hat{H})$ or minimizes $E$ for a fixed entropy $S_{vn}$. In this case $S_{vn}=S_H$.
\par
Similarly to the classical Hamiltonian evolution the reversible dynamics given by \eqref{vN_eq} preserves entropy and hence cannot 
describe the equilibration process for an isolated quantum system without additional coarse-graining procedures. 
In particular, a pure state represented in the Hamiltonian eigenbasis by $|\psi\rangle = \sum_j c_j |j\rangle$
remains a pure state. von Neumann proposed as a first step towards thermalization the time-averaging procedure leading from $|\psi\rangle$ 
to the following density matrix (for a generic case of a non-degenerated Hamiltonian spectrum).
\begin{equation} 
\hat{\rho}_{D} =  \lim_{\tau\to\infty} \frac{1}{\tau}\int_0^{\tau}  e^{-i/\hbar\hat{H} t} |\psi\rangle\langle\psi| e^{-i/\hbar\hat{H} t} d\tau
= \sum_j |c_j|^2  |j\rangle\langle j| .
\label{ergodic_av}
\end{equation}
The problem of thermalization mechanism for closed,  complex quantum system is still open \cite{rigol2008thermalization}.

\subsection{Finite quantum systems}

To avoid mathematical problems we begin with the discussion of equilibrium states for quantum systems with finite-dimensional Hilbert spaces.
The basic property of an equilibrium system is related to the Kelvin formulation of the Second Law: \emph{It is not possible to extract work from a single heat source at a fixed temperature in a cyclic process}\cite{callen1998}. This leads to the notion of a \emph{passive state} \cite{haag1974,lenard1978,pusz1978passive}
 for a given system with a Hamiltonian $\hat{H}$ as the state $\hat{\rho}$ for which 
\begin{equation} 
 \mathrm{Tr}(\hat{\rho} \hat{H}) \leq  \mathrm{Tr}(\hat{U}\hat{\rho}\hat{U}^{\dagger} \hat{H}) 
\label{passive_df}
\end{equation} 
for any unitary $\hat{U}$. This arbitrary unitary map represents any reversible  external driving applied to the system and the inequality \eqref{passive_df} 
means impossibility of extracting work by such a procedure. It is not difficult to show that any passive state 
$\hat{\rho}_p$ is diagonal in the Hamiltonian eigenbasis which can be ordered in such a way that
\begin{equation} 
\hat{\rho}_p = \sum_{j=1}^n \lambda_j  |j\rangle\langle j|, \quad   E_j \leq E_{j+1}, \quad \lambda_{j+1} \leq \lambda_j
\label{passive_state}
\end{equation}
where $\hat{H} |j\rangle = E_j |j\rangle$.
\par
Gibbs states \eqref{gibbs} are obviously passive, but there exist many others, like for instance a variant of microcanonical ensemble determined by the energy scale $E$ and defined as
\begin{equation} 
\hat{\rho}[E] = \frac{1}{\#\{j; E_j \leq E\}}\sum_{\{j; E_j \leq E\}} |j\rangle\langle j| .
\label{micro_state}
\end{equation}
However, only Gibbs states possess the property of \emph{complete passivity} which means that also its  $n$-fold product $\hat{\rho}^{\otimes n}$ is passive with respect to $n$-fold sum of its Hamiltonian, for arbitrary $n=1,2,3,...$.
No energy can be extracted by a unitary even from the $n$-fold product completely passive state, 
which is a quantum version of Kelvin's II-law.

\par
Any density matrix $\hat{\rho}$ can be transformed into a unique passive state $\hat{\rho}_p = \hat{U} \hat{\rho} \hat{U}^{\dagger}$ by a unitary $\hat{U}$ which maps the eigenvectors of $\hat{\rho}$ into the  eigenvectors of $\hat{H}$ with the proper ordering. 
\par
Kubo introduced multi-time correlation functions (called Green functions) at the equilibrium states as a link between quantum statistical 
mechanics and nonequilibrium dynamics \cite{kubo1957}. Generalizing an idea by Einstein on the relation between drag and restoring force of a brownian particle
Green and Kubo   \cite{green1954markoff} expressed the transport coefficients   in terms of integrals of equilibrium  time correlation functions.
As an illustration consider a two-point correlation function for finite system at the Gibbs state corresponding to the Hamiltonian $\hat{H}$
\begin{equation} 
F_{AB}(t) =  \mathrm{Tr}\bigl(\hat{\rho}_{\beta}\hat{A}(t) \hat{B}\bigr), \quad  \hat{A}(t) = e^{\frac{i}{\hbar}\hat{H}t}\hat{A}e^{-\frac{i}{\hbar}\hat{H}t}
\label{Corr}
\end{equation}
for two observables $\hat{A}$ and $\hat{B}$. Discreteness of the Hamiltonian spectrum implies that $F_{AB}(t)$ is a \emph{quasi-periodic function}, i.e. after sufficient time its value returns arbitrarily close to the initial one, which corresponds to Poincare recurrences in classical mechanics. 
\par
By analytic continuation the functions $F_{AB}(t)$  can be extended to a complex domain ($t\to z$) and one can show that Gibbs states are completely characterized by the following the Kubo-Martin-Schwinger (KMS) condition \cite{kubo1957,martin1959}
\begin{equation} 
F_{AB}(-t) = F_{BA}(t - i\hbar\beta)   
\label{KMS}
\end{equation}
valid for any pair of observables and arbitrary time.

\subsection{Infinite quantum systems and KMS states}

Large, many-particle quantum systems are important in quantum thermodynamics for studying  physical properties  
of bulk matter or models of heat baths in the context of nonequilibrium theory of open systems. 
A very useful idealization called the \emph{thermodynamic limit} is a mathematical procedure replacing 
a system  of $N$ particles in a volume $V$ by its infinite volume limit with a fixed density $N/V$. 
The mathematically rigorous theory of infinite quantum systems has been developed in the 60-ties and 70-ties and allowed to study, 
for example, decay of spatial and temporal correlations or define precisely the notion of phase transition and spontaneous symmetry breaking.
\par
The original Hilbert space description in terms of density matrices and hermitian operators loses its meaning in the thermodynamical limit 
and must be replaced by a more abstract algebraic formalism. However, 
one can use an alternative approach involving Green functions, for which their thermodynamic limit can be  well-defined.
\par
As an example, consider a free Bose or Fermi gas confined in a finite volume and described by a set of annihilation and creation operators
$\hat{a}_k $ , $\hat{a}^{\dagger}_k $ labeled a discrete set of quantum numbers $\{ k\}$ 
and satisfying canonical commutation and anticommutation relations, respectively. The Hamiltonian is given by
\begin{equation} 
\hat{H} = \sum_{k} \epsilon_k \hat{a}^{\dagger}_k \hat{a}_k ,
\label{Ham_gas}
\end{equation}
and the thermal equilibrium state  by the \emph{grand canonical ensemble}
\begin{equation} 
\hat{\rho}_{\beta,\mu} =  Z^{-1}(\beta,\mu) e^{-\beta \hat{H}_{\mu}} , \quad Z(\beta,\mu) = \mathrm{Tr}e^{-\beta \hat{H}_{\mu}}
\label{gibbs_GCE}
\end{equation}
which can be treated as a Gibbs state with the modified Hamiltonian 
\begin{equation} 
\hat{H}_{\mu} = \hat{H}-\mu\hat{N} = \sum_{k} (\epsilon_k -\mu) \hat{a}^{\dagger}_k \hat{a}_k .
\label{H_mod}
\end{equation}
where $\mu$ is a chemical potential.
\par
For a pair of observables $A = \sum_k (f_k \hat{a}_k  + \bar{f_k} \hat{a}^{\dagger}_k)$ and $B = \sum_k (g_k \hat{a}_k  
+ \bar{g_k} \hat{a}^{\dagger}_k)$ the Green function in the thermodynamic limit can be computed replacing the discrete energy levels
$\epsilon_k$ by the continuous variable  $\hbar|\omega|$ and $f_k$, $g_k$ by functions $f(|\omega|, \alpha)$, $g(|\omega|, \alpha)$ where $\alpha$ denotes additional (discrete and continuous) quantum numbers. Then
\begin{eqnarray} 
F_{AB}(t) &=&  \int_{-\infty}^{+\infty}d\omega\, e^{-i\omega t}\int d\alpha\, \bigl\{\bar{f}(|\omega|, \alpha)g(|\omega|, \alpha) [1 -(\mp) n(\hbar|\omega|)]\Theta(\omega)\nonumber\\
&+&\bigl[f(|\omega|, \alpha)\bar{g}(|\omega|, \alpha)  n(\hbar|\omega|)\Theta(-\omega)\big\}, 
\label{Corr_limit}
\end{eqnarray}
where $\int d\alpha$ denotes integral and sum over continuous or discrete $\alpha$-s , $\Theta(\cdot)$ is the Heaviside function and
\begin{equation} 
n(x) = \frac{1}{e^{\beta (x - \mu)}\mp 1} ,
\label{population}
\end{equation}
with the convention that in $\mp$ the minus sign corresponds to bosons and the plus sign to fermions. The Green function \eqref{Corr_limit} has an explicit structure of a Fourier transform that illustrates the fact that in the thermodynamic limit time correlations decay to zero for long times without Poincare recurrences. Moreover, one can expect that in the generic case of infinite systems the inverse Fourier transforms $\tilde{F}_{AB}(\omega)$ are meaningful and then the KMS condition \eqref{KMS} implies the relation \cite{kubo1957,martin1959}
\begin{equation} 
\tilde{F}_{BA}(-\omega) = e^{-\hbar\beta\omega}\tilde{F}_{AB}(\omega),
\label{KMS_1}
\end{equation}
which plays an important role in the quantum theory of open systems.
\par
The KMS condition in the form \eqref{KMS} has been proposed to define thermal equilibrium states for infinite systems \cite{kubo1957,martin1959}.
It has been subsequently proved that KMS states possess desired stability properties with respect to local perturbations. 
Moreover, passivity (originally introduced in context of infinite systems \cite{bratteli1996operator}) combined with a certain clustering property, which excludes long-range order, implies the KMS condition \cite{pusz1978passive}. 
\par
For finite systems at the  given temperature the corresponding Gibbs state is unique. In the case of an infinite system at the given temperature many KMS states can coexist, usually below a certain critical temperature. This is exactly the mechanism of \emph{phase transition}, the notion which can be precisely defined only in the thermodynamic limit  
\chpots.

\section{Non-equilibrium Thermodynamics of Quantum Open Systems}

The progress in the field of quantum optics and laser physics in 60-ties and 70-ties stimulated  efforts to develop a mathematically sound theory of irreversible quantum dynamics. 
As noticed by Kraus \cite{Kraus71}, the mathematical theory of completely positive (CP) maps \cite{stinespring1955} provided a natural framework for both, the dynamics of open quantum systems and quantum measurement theory. The general form of CP and trace preserving map reads
\begin{equation} 
\Lambda{\hat\rho} = \sum_{j} \hat{W}_j^{\dagger} \hat{\rho} \hat{W}_j,
\label{Kraus}
\end{equation}
where $\hat{W}_j$ are called \emph{Kraus operators} and satisfy the condition $\sum_{j} \hat{W}_j\hat{W}_j^{\dagger} = I$.
\par
For any CP dynamical map $\Lambda$, Lindblad proved a kind of \emph{H-theorem} \cite{lindblad75}
\begin{equation} 
S(\Lambda\hat{\rho}|\Lambda\hat{\sigma}) \leq S(\hat{\rho}|\hat{\sigma})
\label{Hth}
\end{equation}
valid for the \emph{relative entropy} of an arbitrary pair of density matrices 
\begin{equation} 
S(\hat{\rho}|\hat{\sigma}) = \mathrm{Tr}\bigl(\hat{\rho}\ln\hat{\rho} - \hat{\rho}\ln\hat{\sigma}\bigr) .
\label{relent}
\end{equation}
The highlight of this period was the discovery in 1976 of the general form of the Markovian Master Equation (MME) satisfying CP condition 
\begin{equation} 
\frac{d}{dt}{\hat\rho} = -\frac{i}{\hbar}[\hat{H} , \hat{\rho}] + \frac{1}{2}\sum_{j} ([\hat{V}_j \hat{\rho}, \hat{V}_j^{\dagger}]+[\hat{V}_j ,\hat{\rho} \hat{V}_j^{\dagger}])\equiv-\frac{i}{\hbar}[\hat{H} , \hat{\rho}] +\mathcal{L}\hat{\rho}\, .
\label{GKLS}
\end{equation}
called the Gorini-Kossakowski-Lindblad-Sudarshan (GKLS) equation \cite{chruscinski2017brief}.
\par
While in \cite{gorini1976completely}  finite-dimensional Hilbert spaces were considered, the case of bounded generators $\mathcal{L}$ for open systems with infinite-dimensional spectrum was independently proved in \cite{lindblad76}. For a recent discussion of the still open unbounded case see \cite{siemon2017unbounded}.

\subsection{Quantum Thermodynamics in the Markovian regime}

Two years before the appearance of the GKLS equation Davies presented a rigorous derivation of MME for a $N$-level system weakly coupled to a heat bath represented by a an ideal fermionic gas at the thermodynamic limit \cite{davies74}. The derivation incorporates in a single mathematical procedure, called weak coupling limit (WCL), which includes the heuristic ideas of Born, Markovian and secular approximations, previously applied to various examples of open systems such as nuclear magnetic resonance by Bloch \cite{wangsness1953dynamical} and later Redfield \cite{redfield1957theory}. Other approaches to the MME include the projection technique of Nakajima-Zwanzig  \cite{nakajima1958quantum,zwanzig1960ensemble}.
\par
Adding to the WCL method a kind of renormalization procedure which allows to  use the physical Hamiltonian $\hat{H}$ of the system, containing lowest order \emph{}
Lamb corrections, and parametrizing the system-bath interaction as $\hat{H}_{int} = \sum _k \hat{S}_k\otimes\hat{R}_k $  one obtains the following structure of MME
which is in the GKLS form
\begin{equation}
\frac{d}{dt}{\hat\rho} = -i[\hat{H} , \hat{\rho}] + \mathcal{L}\hat{\rho}, \quad  \mathcal{L}\hat{\rho}=\sum_{k,l}\sum_{\{\omega
\}}\mathcal{L}_{lk}^{\omega}\hat{\rho}
\label{Davies}
\end{equation}
where 
\begin{equation}
\mathcal{L}_{lk}^{\omega}\hat{\rho} ={\frac{1}{2\hbar^2}}\tilde{R}_{kl}(\omega )\left\{ [\hat{S}_{l}(\omega )\hat{\rho}
,\hat{S}^{\dagger }_{k}(\omega )]+[\hat{S}_{l}(\omega ),\hat{\rho} \hat{S}^{\dagger }_{k}(\omega
)]\right\} .  
\label{Davies1}
\end{equation}
Here, the operators $\hat{S}_{k}(\omega)$ originate from the Fourier decomposition ($\{\omega\}$- denotes the set of Bohr frequencies of $\hat{H}$).
\begin{equation}
e^{i/\hbar\hat{H} t}\hat{S}_{k}e^{-i/\hbar\hat{H} t}= \sum_{\{\omega\}} e^{-i\omega t}\hat{S}_k(\omega ) ,
\label{FourierS}
\end{equation}
and $\tilde{R}_{kl}(\omega )$ is the  Fourier transform of the bath correlation function $\langle \hat{R}_k(t) \hat{R}_l\rangle_{bath}$ computed in the thermodynamic limit $\tilde{R}_{kl}(\omega ) = \int_{-\infty}^{+\infty} e^{i\omega t} \langle \hat{R}_k (t) \hat{R}_l \rangle_{bath} dt$. 
The derivation of  \eqref{Davies},\eqref{Davies1} makes sense for a generic stationary state of the bath and implies two properties:\\
1) the Hamiltonian part $[\hat{H}, \cdot]$ commutes with the dissipative part $\mathcal{L}$,\\
2) the diagonal (in $\hat{H}$-basis) matrix elements of $\hat \rho$ evolve (independently of the off-diagonal ones) according to the Pauli Master Equation with transition rates given  by the Fermi Golden Rule \cite{fermi1950nuclear,alicki1977markov}.\\
If additionaly the bath is a heat bath, i.e. an infinite system in a KMS state the additional relation \eqref{KMS} implies that:\\
3) Gibbs state $\hat{\rho}_{\beta} = Z^{-1} \exp{-\beta\hat{H}}$ is a stationary solution of \eqref{Davies},\\
4) under the condition that only scalar operators commute with all $\{\hat{S}_{k}(\omega), \hat{S}^{\dagger}_{k}(\omega)\}$, any initial state relaxes asymptotically to the Gibbs state:
{\em The 0-Law of Thermodynamics} \cite{frigerio1977quantum}.

The derivation of \eqref{Davies},\eqref{Davies1} can be extended to slowly varying time-dependent Hamiltonian (within the range of validity of the adiabatic theorem)  \cite{davies1978} $H(t)$ and an open system coupled to several heat baths at the inverse temperatures $\{\beta_k= 1/k_B T_k\}$. The MME takes form 
\begin{equation}
\frac{d}{dt}\hat{\rho}(t)  = -i[\hat{H}(t),\hat{\rho}(t)] + \mathcal{L}(t)\hat{\rho}(t),\quad \mathcal{L}(t) = \sum_k \mathcal{L}_k(t) .
\label{master2}
\end{equation}
Each  $\mathcal{L}_k(t)$ is derived using a temporal Hamiltonian $\hat{H}(t)$, $\mathcal{L}_k(t) \hat{\rho}_j(t)=0$ with a temporary Gibbs state $\hat{\rho}_j(t) = Z_j^{-1}(t) \exp\{-\beta_j \hat{H}(t)\}$. The energy conservation in this case is the  First Law of Thermodynamics \cite{alicki1979quantum}
\begin{equation}
\frac{d}{dt}E(t) = {\cal J}(t) -  {\cal P}(t) .
\label{work_heat}
\end{equation}
Here 
\begin{equation}
E(t)= \mathrm{Tr}\Bigl(\hat{\rho}(t)\hat{H}(t)\Bigr)
\label{internal}
\end{equation}
is the internal energy of the system, 
\begin{equation}
{\cal P}(t) \equiv -\mathrm{Tr}\Bigl(\hat{\rho}(t) \frac{d \hat{H}(t)}{dt}\Bigr),\quad 
\label{power}
\end{equation}
is the power  provided by the system, and 
\begin{equation}
{\cal J}(t) \equiv \mathrm{Tr}\Bigl(\hat{H}(t) \frac{d}{dt}\hat{\rho}(t)\Bigr) =\sum_k {\cal J}_k(t) , \quad {\cal J}_k(t) = \mathrm{Tr}\Bigl(\hat{H}(t) \mathcal{L}_k(t)\hat{\rho}(t)\Bigr) .
\label{heat}
\end{equation}
is the sum of net heat currents supplied by the individual heat baths.

The $H$- theorem \eqref{Hth} directly implies the following mathematical identity \cite{spohn1978entropy} and  \cite{mcadory1977}\
\begin{equation}
- \mathrm{Tr} \left[ \mathcal L \hat \rho(t) \left( \ln \hat \rho(t) - \ln \hat {\rho}_{st} \right) \right] \geq 0 ~, \mathrm{for} \quad \mathcal{L}\hat {\rho_{st}} = 0 ~,
\label{eq:spohn}
\end{equation}
which applied to individual generators $\mathcal{L}_k(t)$ reproduces the Second Law of Thermodynamics in the form 
\begin{equation}
\frac{d}{dt} S_{vn}(t) - \sum_k \frac{1}{T_k} {\cal J}_k(t) \geq 0 .
\label{IIlaw}
\end{equation}
obtained first for the constant $\hat{H}$ in \cite{spohn1978irreversible} and  ultimately generalized in \cite{alicki1979quantum}.
\par
For external periodic modulation of the Hamiltonian $\hat{H}(t) = \hat{H}(t+\tau) $, 
a very similar WCL formalism for open systems has been developed    \cite{k114,kohler1997floquet,k275,szczygielski2013markovian}. 
One assumes that modulation is fast, i.e. its angular frequency $\Omega = 2\pi/\tau$ is comparable to the relevant Bohr frequencies of the Hamiltonian, therefore the previous adiabatic approximation is not appropriate. According to the Floquet theory the unitary propagator \eqref{vN_solution} $\hat{U}(t) \equiv \hat{U}(t,0)$ can be written as 
\begin{equation}
\hat{U}(t) = \hat{U}_p(t) e^{-\frac{i}{\hbar} \hat{H}_{av}t}
\label{Floquet}
\end{equation}
where $\hat{U}_p(t)= \hat{U}_p(t+\tau)$ is a periodic propagator and $\hat{H}_{av}$ can be called \emph{averaged Hamiltonian}. Under similar assumptions as before one can derive, 
using the WCL procedure, the Floquet- Markovian ME in the \emph{interaction picture}

\begin{equation}
\frac{d}{dt}{\hat\rho}^{int}(t) = \mathcal{L}\hat{\rho}^{int}(t), \quad  \mathcal{L}\hat{\rho}=\sum_{k,l}\sum_{\{\omega_q
\}}\mathcal{L}_{lk}^{\omega_q}\hat{\rho}
\label{Floq_Dav}
\end{equation}
where 
\begin{equation}
\mathcal{L}_{lk}^{\omega_q}\hat{\rho} ={\frac{1}{2\hbar^2}}\tilde{R}_{kl}(\omega_q )\left\{ [\hat{S}_{l}(\omega_q )\hat{\rho}
,\hat{S}^{\dagger }_{k}(\omega_q )]+[\hat{S}_{l}(\omega_q ),\hat{\rho} \hat{S}^{\dagger }_{k}(\omega_q
)]\right\} .  
\label{Floq_Dav1}
\end{equation}
Now, the summation in \eqref{Floq_Dav} is taken over the set of \emph{extended Born frequencies}\,  
$\{\omega_q = \omega_{av} + q\Omega | \omega_{av} - \mathrm{Bohr\ frequencies\ of} \hat{H}_{av}, q\in \mathbf{Z}\}$, 
which takes into account the exchange processes of energy quanta $\hbar|q|\Omega$ with the source of external modulation. Here again the operators $\hat{S}_{k}(\omega_q)$ originate from the Fourier decomposition 
\begin{equation}
\hat{U}^{\dagger}(t) \hat{S}_{k} \hat{U}(t)= \sum_{\{\omega_q\}} e^{-i\omega_q t}\hat{S}_k(\omega_q ) .
\label{FourierS}
\end{equation}
Notice, that the interaction picture generator is time-independent and  the Schr\"odinger picture dynamics is
given by the composition $\hat{\rho} \mapsto  \hat{U}(t) \bigl(e^{\mathcal{L}t}\hat{\rho}\bigr)\hat{U}^{\dagger}(t)$. Typically, $\mathcal{L}$ possesses a single stationary state $\hat{\rho}_0$ and then for any initial state $\hat{\rho}(0)$ the Schr\"odinger evolution drives the system to a limit cycle  
$\hat{\rho}_{lc}(t) = \hat{U}_p(t)\hat{\rho}_0\hat{U}_p^{\dagger}(t)$.
\par
Heat currents corresponding to different baths can be defined for any time. 
As a result the Second Law is satisfied for this definition, nevertheless the form of the First Law is problematic. 
Namely, for fast modulation the instantaneous decomposition of  energy into work and internal energy of the system is not clear. 
Only in the limit cycle, where the system's internal energy and entropy are constant, and the heat currents are time independent we can write the First Law as
\begin{equation}
{\cal P} = \sum_j {\cal J}_j ,
\label{power_per}
\end{equation}
and the Second Law as
\begin{equation}
\sum_j \frac{1}{T_j} {\cal J}_j \leq 0 .
\label{IIlaw_per}
\end{equation}
Here, the heat current associated with the $j$ bath is given in terms of the corresponding interaction picture generator
\begin{equation}
{\cal J}_j = \sum_{l,k \in I_j}\sum_{\{\omega_q\}}\frac{\omega_q}{\omega_{av}}\mathrm{Tr}\bigl(\hat{H}_{av}\mathcal{L}_{lk}^{\omega_q}\hat{\rho}_0\bigr) ,
\label{Floq_Dav1}
\end{equation}
and $I_j$ denotes the subset of indices corresponding to the interaction with the $j$-th heat bath.
The above scheme has been extended to non-equilibrium stationary baths in \cite{alicki2015non}, 
with possible applications to non-thermal radiation baths, rotating heat baths, etc. \cite{rossnagel2014nanoscale,niedenzu2017universal}.

\subsection{Beyond the WCL Markovian Approximation}

The theory of open quantum systems together with the Davies construction supplies a consistent framework of Thermodynamics 
where the basic laws have a quantum dynamical  framework \cite{k281}. This framework is quite restrictive and therefore one may ask if some
of the assumptions can be relaxed without compromising the consistency with thermodynamics.

Many suggestions have been proposed:
\begin{itemize}
\item{Challenging complete positivity.}
\item{Local vs Non Local GKLS equation.}
\item{Non Markovian dynamics.}
\item{Strong system-bath coupling.}
\end{itemize}

The complete positivity structure assumes that initially the system and bath are uncorrelated \cite{Kraus71}. 
This has been challenged by Pechukas \cite{pechukas1994reduced} who claimed that positivity of the dynamical map is sufficient.
Alicki  responded that one should stress that beyond the weak
coupling regime there exists no unique definition of the quantum reduced dynamics \cite{alicki1995comment}. 
A similar answer was given  by Lindblad  \cite{lindblad1996existence}.
It has been claimed that the second law of thermodynamics is violated by a non-CP dynamics \cite{argentieri2014violations,argentieri2015complete}.

An alternative approach to open system dynamics has been proposed by Caldeira,  and Leggett, based on a path integrals formalism,
generating a QME for quantum Brownian motion \cite{caldeira1983path}. The equation is not guaranteed to be positive in particular at low temperature.
A fix to the problem has been suggested by Diosi adding terms to the equation to obtain a GKLS format \cite{diosi1993calderia}.
For a Brownian particle one would expect that the friction is isotropic meaning that the dissipation equations should be translational invariant.
It has been noticed by Tannor and Kohen that complete positivity, translation invariance and detailed balance cannot be satisfied simultaneously
\cite{kohen1997phase,lindblad1976brownian}. This is also true for the fix of Diosi which adds a diffusion-like term in position.

The Davies construction of the GKLS equation Eq. (\ref{Davies}) requires that the jump operators Eq. (\ref{Davies1})
are generated from the complete system Hamiltonian Eq. (\ref{FourierS}).  What happens when the system can be deconstructed into  segments
which are weakly coupled to each other? Can one use a local GKLS equation for each segment and then linking together to construct a network?
Careful analysis has shown that such a construction can violate the II-law: Heat can flow from the cold to the hot bath spontaneously \cite{k290}.
In degenerate networks when the links are identical the secular approximation may fail for vanishing small links. In these cases local GKLS equations
give the correct heat current with respect to numerical converged approaches \cite{hofer2017markovian,mitchison2017non,gonzalez2017testing}.
General conditions of adding up consistently GKLS generators have been suggested \cite{brask2017additivity}.

A violation of the II-law has also been identified if the Floquet GKLS equation Eq. (\ref{Floq_Dav1}) is replaced by the standard stationary GKLS.
This is even true for the well known two-level Bloch equation \cite{k114} and for the three-level amplifier \cite{k122}.

It is customary to start the non-Markovaian investigation from the second order integro-differential equation 
\cite{nakajima1958quantum,zwanzig1960ensemble,haake1973statistical,kleinekathofer2004non,de2017dynamics}:
\begin{equation}
\frac{d}{dt} \hat \rho_s = -i [\hat H_{eff} ,\hat \rho_s ] + \int_0^t ~ dt'{\cal K}(t,t') \hat \rho_s(t')
\label{integro}
\end{equation}
where $\hat H_{eff}$ is an effective system Hamiltonian and ${\cal K}(t,t')$ is termed a memory kernel.

Different approaches can be classified by the type of approximation to the memory kernel.
One option is to generate a time local kernel \cite{shibata1977generalized,kofman2004unified} which can lead to a GKLS-like equations
with time dependent coefficients. Complete positivity is not ensured which manifests itself by negative coefficients. 

Another option is termed the Hierarchical Equations of Motion Approach \cite{tanimura1989time,meier1999non,kleinekathofer2004non,jin2008exact,ishizaki2009unified}
which decomposes the kernel to exponentially decaying functions \chheom. One then adds a set of auxiliary variables which leads to a hierarchy of coupled
differential equations. This is equivalent to a Markovian description embedded in a larger Hilbert space.
The thermodynamical consequence of non-Markovian dynamics has recently been addressed \cite{marcantoni2017entropy}.
It has been observed that in the absence of the semigroup property, if the reduced dynamics has a thermal asymptotic
state, this need not be stationary. Then even the integrated entropy production becomes negative.
These observations imply that, when the conditions leading to reduced dynamics of semigroup type are
relaxed, a consistent formulation of the second law of thermodynamics requires that the environment
contribution to the entropy balance be explicitly taken into account \cite{alipour2016correlations}.

An alternative  theory of quantum thermodynamics in the framework of the nonequilibrium
Green's functions has been proposed by Esposito and Galperin \cite{esposito2015quantum,ludovico2016dynamics}. The theory was applied to noninteracting
open quantum systems strongly coupled to their reservoirs. The theory is non-Markovian and nonlocal in time.
As a consequence the particle number, energy, and entropy of the system are redefined
as energy-resolved versions of the standard weak coupling definitions. The approach has been criticised as 
failing, already at equilibrium,  to describe correctly the energy fluctuations \cite{ochoa2016energy}.

Strong system bath coupling is another challenge that has been met by embedding in a larger system. 
The main idea is to move the system bath partition further into the bath \chreacc.
The polaron transformation is such an example.
It incorporates part of the bath degrees of freedom in a modified system 
\cite{schaller2013single,segal2014two,xu2016polaron,wang2017unifying,gelbwaser2015strongly,bruch2016quantum,strasberg2016nonequilibrium,newman2017performance,perarnau2018strong}.
If weak coupling is incorporated on the new system bath boundary, consistency with thermodynamics is maintained.

Another approach to strong coupling is to embed the system in a finite surrogate spin bath which represents the true infinite bath. 
The total system and bath are described by unitary dynamics.
To model the infinite bath thermal boundary conditions are imposed between a thermal
secondary bath and the primary bath. A random swap operation is employed for this task. 
Each individual realization is unitary. Averaging the individual realizations  is equivalent to a
Poisson type GKLS equation on the boundary of the primary bath. 
Thermodynamic properties can be obtained by evaluating the currents through the device \cite{k307}.
Consistency with thermodynamics has been obtained for the case of heat transfer 
from a hot to a cold bath irrespective of the system-bath coupling \cite{k307,k306}.

\section{Models of Quantum Engines and Refrigerators}

Since the pioneering work of Carnot \cite{carnot1872}, learning from example has been a major theme in thermodynamical studies. 
This is also true in QT where the issues of heat and work obtain a concrete meaning \cite{alicki1979quantum}.
In addition the tradeoff between efficiency and finite power can be explored.
The trend toward miniaturisation has led to the construction of quantum heat devices composed from a microscopic working entity, 
for example a single ion in a Paul trap \cite{Rossnagel325} ~\chsion.
This macroscopic scale raises the question: What quantum effects to expect? Is there a role for coherence or entanglement? 
Can we expect quantum supremacy? 

Models of heat engines and refrigerators can lead to new insight in QT. They can be broadly classified as reciprocating and continuous.

\subsection{Reciprocating engines and refrigerators}

Reciprocating engines are composed of a series of strokes which combine to a cyclic operation.
The different cycles are defined by the individual stroke operations and their order. 
In QT a reciprocating engine can be defined by a product of CP maps Eq. (\ref{Kraus}), which operate on the working medium:
\begin{equation}
{\cal U}_{cyc} = \prod_j {\cal U}_j
\label{recip-1}
\end{equation}
where ${\cal U}_{cyc}$ is the cycle propagator and ${\cal U}_j$ are stroke propagators. The steady state operation
is an invariant of the cycle propagator ${\cal U}_{cyc} \hat \rho_{st}=\hat \rho_{st}$.
For cycles that have a single non-degenerate invariant the CP character of ${\cal U}_{cyc}$, Eq. (\ref{relent})
guarantees a monotonic convergence to the steady state cycle, termed the limit cycle \cite{k201}.

The four stroke Otto cycle is a primary example \chpqtc  ~\chsaqt. It is composed of two unitary strokes and two thermalization strokes:
The Hamiltonian of the working medium is parametrically externally controlled: $\hat H (\omega)$ where 
$\omega$ is an external parameter which changes the energy scale.
For example $\hat H = \frac{1}{2m}\hat P^2+ \frac{m \omega(t)^2}{2} \hat X^2$ for the harmonic working medium \cite{k221} 
and $\hat H = \omega(t) \hat S_z + J \hat S_x$ for a spin system \cite{k85,k190}.

The quantum  Otto cycle is therefore described as:
\begin{enumerate}
\item{The hot {\em isochore}: heat is transferred from the hot bath to the working medium without  change in the external parameter $\omega_h$.
The stroke is described by the propagator ${\cal U}_h$.}
\item{The expansion {\em adiabat}: the working medium reduces its energy scale from $\omega_h $ to $\omega_c$, 
with $\omega_h > \omega_c$, producing work while isolated from the hot and cold reservoirs. The stroke is described by the propagator ${\cal U}_{hc}$.}
\item{The cold {\em isochore}: heat is transferred from the working medium to the cold bath without change in the external parameter $\omega_c$.
The stroke is described by the propagator ${\cal U}_c$.}
\item{The compression {\em adiabat}: the working medium increases its energy scale from $\omega_c $ to $\omega_h$,
consuming power while isolated from the hot and cold reservoirs. The stroke is described by the propagator ${\cal U}_{ch}$.}
\end{enumerate}
The cycle propagator becomes the product of the segment propagators:
\begin{equation}
 {\cal U}_{cyc}={\cal U}_{ch}{\cal U}_c {\cal U}_{hc}{\cal U}_h\;.
 \label{eq:cycprop}
 \end{equation}
 It should be mentioned that the stroke propagators do not commute for example:
 $[{\cal U}_{hc},{\cal U}_h] \ne 0$. The Otto cycle can operate in two extreme protocols, adiabatic and sudden. 
 
In the adiabatic cycle the working medium state is diagonal in the energy representation throughout the cycle. 
Such cycles are called stochastic \cite{esposito2012stochastic,seifert2012stochastic}. 
The efficiency becomes $\eta_o= 1-\frac{\omega_c}{\omega_h} \le \eta_c$  where  $\eta_c=1-\frac{T_c}{T_h}$  is the Carnot efficiency. 

To obtain finite power the time allocated to the propagators ${\cal U}_{hc}$ and ${\cal U}_{ch}$ should be shortened. 
For the unitary strokes this means deviating from the adiabatic limit.
Whenever $[\hat H (t),\hat H(t')]\ne 0$ coherence will be generated and $\hat \rho$ will not be diagonal in the energy basis $S_H > S_{vn}$.
Generating coherence will always cost additional external work. This phenomena has been termed quantum friction \cite{k190,plastina2014}.
Quantum friction can be understood using the notion of passivity Eq. (\ref{passive_df}). 
In the adiabatic limit the eigenvalues of the density operator remain passive in the energy basis. 
The minimum work  can be associated to the change in energy scale. 
Any nonadiabatic deviation will increase the required work.

The price of generating coherence can be reduced if at the end of the adiabatic stroke the state is restored to be 
passive  in the energy basis. 
Such protocols are termed shortcuts to adiabticity or frictionless \cite{muga09,torrontegui2013shortcuts,del2014more} \chsaqt.
These protocols allow to achieve adiabatic like solutions in finite time for the propagators ${\cal U}_{hc}$ and ${\cal U}_{ch}$.
The fast shortcut solutions raise the question what is the shortest time allocation for frictionless adiabatic strokes.
This issue is in the realm of the quantum speed limit \cite{anandan1990geometry,deffner2013energy} 
with the caveat that the energy scale of the Hamiltonian also changes.
The transformation can be made faster if temporary energy is stored in the working fluid. 
Optimal control protocols that constrain the stored energy in the working fluid
lead to a scaling of the time allocation as $\tau \propto \frac{1}{\sqrt{\omega_c\omega_h}}$ for $\omega_c \rightarrow 0$ \cite{chen2010fast,stefanatos2017minimum}.

Coherence can also be introduced as a resource by employing a non-thermal bath. Even single bath is sufficient to extract work \cite{scully2003extracting}.
Nevertheless there is  no violation of the II-law if accounting is done properly \cite{niedenzu2016operation}.

For finite power also the time allocated to thermalization ${\cal U}_{c}$ and ${\cal U}_{h}$ should be restricted. 
Typically in most studies the generator of thermalization ${\cal L}$ is the GKLS equation (\ref{GKLS}) \cite{k85}.
Finite time allocation is obtained  by avoiding the infinite time full thermalization.
Optimizing the time allocation in the stochastic limit leads to a finite power engine. 
The efficiency at maximum power at high temperature becomes \cite{k221}:
\begin{equation}
\eta_{ca} = 1 -\sqrt{\frac{T_c}{T_h}}
\label{ca-efficiency}
\end{equation}
which is known as the Novikov-Curzon-Ahlborn efficiency \cite{novikov1958efficiency,curzon75}.
The importance of Equation (\ref{ca-efficiency}) is  that it points to the tradeoff between efficiency and power.
For the Otto cycle at high temperature the efficiency at maximum power is limited by the energy level structure of the working medium with the leading term $\eta \approx \frac{1}{2} \eta_c + ..$ \cite{k294}.
This result has been obtained from general considerations in the adiabatic limit \cite{cavina2017slow}.

In QT the Carnot cycle has received less attention than the Otto cycle. The reason is that  the hot and cold isochores are replaced by isotherms where the thermalization takes place with  a time dependent Hamiltonian. 
In the adiabtic limit of slow change GKLS equations of motion can be obtained \cite{k85}.
Beyond the adiabatic limit deriving GKLS equations is complicated due to the non-periodic driving. 
The original motivation for the study of QT cycles was to supply a more fundamental justification 
for the empirical Finite-Time-Thermodynamics approach \cite{andresen1977,salamon01}.
In the  infinitely slow cycle limit,  the efficiency converges to the ideal Carnot  efficiency $\eta_c$. 
Optimizing power leads to the Novikov-Curzon-Ahlborn efficiency $\eta_{ca}$, 
which is universal in the stochastic low dissipation limit \cite{k87,esposito2010efficiency}.
In this limit the irreversibility can be associated to heat transport, thus termed endo-reversible.

A two stroke engine has been suggested where ${\cal U}_{cyc}={\cal U}_T {\cal U}_S$ \cite{quan2007quantum,allahverdyan2008work}. 
One that resembles the Otto cycle is composed of a four level working medium.
Thermalization, ${\cal U}_T$ is conduced in parallel where two-levels are connected to the hot bath 
and the other two-levels to the cold bath.
The unitary ${\cal U}_S$ stroke is composed of a swap propagator between these two sets of levels. 
The efficiency of this engine is equivalent to the Otto efficiency $\eta_o$.

The other extreme operational limit is the sudden limit where a limited action is performed on each stroke.
The work per cycle then decreases but the power which is the work divided by cycle time can reach a constant.
In this limit each stroke can be expressed as ${\cal U}_j =\exp{ {\cal L}_j \tau}$, where ${\cal L}_j$ is the generator and $\tau$ the time allocation.
Then  a four stroke cycle becomes equivalent to a continuous engine with finite power \cite{k299} \chqfs .
In the limit of $\tau \rightarrow 0$:
\begin{equation}
 {\cal U}_{cyc}={\cal U}_{ch}{\cal U}_c {\cal U}_{hc}{\cal U}_h=e^{{\cal L}_{ch}\frac{1}{2} \tau}e^{{\cal L}_{c} \tau}e^{{\cal L}_{hc} \tau}e^{{\cal L}_{h} \tau}e^{{\cal L}_{ch}\frac{1}{2} \tau} \approx e^{({\cal L}_{ch}+{\cal L}_{c}+{\cal L}_{hc}+{\cal L}_h)\tau}
 \label{eq:cycprop}
 \end{equation}
which is correct up to $O(\tau^3)$  based on the cyclic property of the engine and Trotter formula \cite{chernoff1974product}.
Moreover the work extraction mechanism employs coherence \cite{k299}. Adding pure dephasing to the engine will null the power
which is a signature of a quantum device. 
\par
Reversing the  sequence of a reciprocating cycle leads to a quantum refrigerator: $ {\cal U}_{cyc}^{ref}={\cal U}_{hc}{\cal U}_c {\cal U}_{ch}{\cal U}_h$.\\
A prerequisite for such a device is that the working medium temperature is lower than the cold bath temperature at the end of the expansion stroke $\omega_h \rightarrow \omega_c$.
Reciprocating refrigeration cycles were used to gain insight on the dynamical approach to the III-law of thermodynamics the vanishing of the cooling power when $T_c \rightarrow 0$ \cite{k152}.
Optimizing the cooling performance requires that the energy gap of the system $\hbar \omega_c$ will match the cold bath temperature $k_B T_c$ \cite{k243}.
The cooling power can either be restricted by the thermalization or by the adiabatic propagator. The energy quant removed from the cold bath per cycle becomes $\hbar \omega_c$.
Considering the optimal frictionless solution, a scaling of ${\cal J}_c \propto T^{\frac{3}{2}}$ is obtained as $T_c \rightarrow 0$.

\subsection{Continuous time quantum machines}

The three-level engine  was the first QT example studied by Scoville et al. \cite{scovil1959,geusic1967quantum}.
The principle of operation is to convert population inversion into output power in the form of  light. 
A hot reservoir characterised by  temperature $T_h$ induces transitions between the ground state
$\epsilon_0$ and the excited state $\epsilon_2$. 
The cold reservoir at temperature $T_c$ couples level $\epsilon_0$ and level $\epsilon_1$.
The amplifier operates by coupling the energy levels $\epsilon_3$ and $\epsilon_2$ to the radiation field  generating an output frequency 
which on resonance is
$\nu = (\epsilon_3-\epsilon_2)/\hbar$. 
The necessary condition for amplification is positive gain or population inversion defined by:
\begin{equation}
{\cal G }= p_2-p_1 \ge 0~.
\label{eq:gain1}
\end{equation}
The positive gain condition dictates:
\begin{equation}
\frac{\omega_c}{\omega_h} \equiv \frac{\omega_{10}}{\omega_{20}} \ge \frac{ T_c}{T_h}~,
\label{eq:ratio}
\end{equation}
The efficiency of the amplifier becomes the Otto efficiency:
$
\eta_o = \frac{\nu}{\omega_{20}} = 1 - \frac{\omega_c}{\omega_h} ~~.
$
Inserting the positive gain condition Eq. (\ref{eq:gain1}) and Eq. (\ref{eq:ratio}) the efficiency is limited by Carnot:
$
 \eta_o ~\le ~\eta_c 
 \label{eq:carnot}
$
This result connecting the efficiency of a quantum amplifier to the Carnot efficiency was first obtained by Scovil et al. \cite{scovil1959,geusic1967quantum}. 
\par
The above description of the 3-level amplifier is based on a static quasi-equilibrium viewpoint.
Real engines which produce power operate far from equilibrium conditions. 
Typically, their performance is restricted by friction, heat transport and heat leaks. 
A  dynamical viewpoint is therefore the next required step \cite{k24}. 
\par
Engines or refrigerators can be classified as either  autonomous or driven.
A  continuous autonomous device operates by connecting to three or more heat baths simultaneously either heating the hottest bath or cooling the coldest bath. 
Such a device operates without any external intervention \cite{k169,tonner2005autonomous,linden2010small}.
A driven system is connected to an external power source or to a more elaborate measurement and feedback device,
which syncronizes the engine \cite{k310}.
\par
The tricycle model is the template for almost all continuous autonomous engines \cite{k272} \chpiqa. 
It can also be viewed as a heat transistor \cite{saira2007heat,schaller2013single}.
Surprisingly very simple models exhibit the same features of engines generating finite power.
Their efficiency at operating conditions is lower than the  Carnot efficiency. 
In addition, heat leaks restrict the performance meaning that reversible operation is unattainable.

\begin{itemize}
\item{The basic model consists of three thermal baths: a hot bath with temperature $T_h$, a cold bath with temperature $T_c$
and a work bath with temperature $T_w$. }
\item{Each bath is connected to the engine via a frequency filter modelled by three oscillators or three qubits:
\begin{equation}
\hat H_F = \hbar \omega_h \hat a^{\dagger} \hat a +\hbar \omega_c \hat b^{\dagger} \hat b + \hbar \omega_w \hat c^{\dagger} \hat c~~,
\label{eq:hfilter}
\end{equation}
where $\omega_h$, $\omega_c$ and $\omega_w$ are the filter frequencies on resonance $\omega_w=\omega_h-\omega_c$.}
\item{The device operates as an engine by removing an excitation from the hot bath and generating excitations
on the cold and work reservoirs. In second quantization formalism  the Hamiltonian describing such an interaction becomes:
\begin{equation}
\hat H_I = \hbar \epsilon \left( \hat a \hat b^{\dagger} \hat c^{\dagger} + \hat a^{\dagger} \hat b \hat c \right)~~,
\label{eq:hinteraction}
\end{equation} 
where $\epsilon$ is the coupling strength. 
}
\item{The device operates as a refrigerator by removing an excitation from the cold bath as well as from the work bath
and generating an excitation in the hot bath. The term $\hat a^{\dagger} \hat b \hat c$ in  the Hamiltonian of Eq. (\ref{eq:hinteraction}) describes this action. }
\end{itemize}
\par
Different types of heat baths can be employed which can include bosonic baths composed of phonons or photons, or
fermonic baths composed of electrons. The frequency filters select from the continuous spectrum of the bath the working
component to be employed in the tricycle. These frequency filters can be constructed also from two-level-systems (TLS) or 
formulated as qubits \cite{k275,skrzypczyk2011smallest,palao13,k289}. A direct realization of Eq. (\ref{eq:hinteraction}) has been perfdormed by 
an absorption refrigerator constructed from three ions in a Paul trap \cite{maslennikov2017quantum}.
\par
The interaction term is strictly non-linear, incorporating three heat currents simultaneously. 
This crucial fact has important consequences. 
A linear device cannot operate as a heat engine or refrigerator \cite{paz2013}. 
A linear device is constructed from a network of harmonic oscillators
with linear connections of the type $\hbar \mu_{ij} \left( \hat a_i \hat a^{\dagger}_j  + \hat a^{\dagger}_i \hat a_j \right)~$ with additional connections
to heat baths constructed from harmonic oscillators. In such a device the hottest bath always cools down and the coldest bath always heats up. 
Thus, this construction
can transport heat but not generate power since power is equivalent to transporting heat to an infinitely hot reservoir. 
Another flaw in a linear model is that the different bath modes do not equilibrate with each other. 
A generic bath should equilibrate any system Hamiltonian irrespective of its frequency.
\par
Many nonlinear interaction Hamiltonians of the type $\hat H_I = \hat A \otimes \hat B \otimes \hat C$ can lead
to a working heat engine. These Hamiltonians can be reduced to the form of Eq. (\ref{eq:hinteraction}) which captures the essence of such interactions.
\par
The first-law of thermodynamics represents the energy balance of heat currents originating from the three baths and collimating on the system:
\begin{equation}
\frac{dE_s}{dt}= {\cal J}_h + {\cal J}_c +{\cal J}_w ~~.
\label{eq:I-law-t}
\end{equation}
At steady state no heat is accumulated in the tricycle, thus $\frac{dE_s}{dt}= 0$. 
In addition, in steady state the entropy is only generated in the baths, leading to the second-law of thermodynamics:
\begin{equation}
\frac{d}{dt}\Delta {\cal S}_u~=~-\frac{{\cal J}_h}{T_h} - \frac{{\cal J}_c}{T_c} -\frac{{\cal J}_w }{T_w}~\ge~0~~.
\label{eq:II-law-t}
\end{equation}
This version of the second-law is a generalisation of the statement of Clausius; heat does not flow spontaneously from cold to hot bodies \cite{clausius1850}.
When the temperature $T_w \rightarrow \infty$, no entropy is generated in the power bath. 
An energy current with no accompanying entropy production  is equivalent to generating pure power:
${\cal P}={\cal J}_w$, where ${\cal P}$ is the output power.
\par
The evaluation of the currents ${\cal J}_j$  in the tricycle model requires dynamical equations of motion. 
A thermodynamical idealisation assumes that the tricycle system and the baths are uncorrelated, meaning that the total state 
of the combined system becomes a tensor product at all times \cite{k281}:
\begin{equation}
\hat \rho = \hat \rho_s \otimes \hat \rho_{H} \otimes \hat \rho_{C} \otimes \hat \rho_{W}~.
\label{eq:rho-t}
\end{equation}
Under these conditions  the  dynamical equations of motion for the tricycle become: 
\begin{equation}
\frac{d}{dt} \hat \rho_s = {\cal L} \hat \rho_s~,
\label{eq:lvn}
\end{equation}
where ${\cal L}$ is  the GKLS Markovian generator Eq. (\ref{Davies}). The derivation of ${\cal L}$ is complicated due to the nonlinearity of the interaction
Eq. (\ref{eq:hinteraction}). Solutions can be obtained in the case when $T_w \rightarrow \infty$  or a singular bath \cite{k272}.
Equivalent solutions for ${\cal L}$ can be obtained for the 3-level \cite{k169}, 2-qubit and 3-qubit absorption refrigerator \cite{correa2014quantum,mu2017qubit}.
\par
The autonomous absorption refrigerator has been a major model in the study of the dynamical version of the III-law of thermodynamics \chcool.
Examining the II-law Eq. (\ref{eq:II-law-t}) as $T_c \rightarrow 0$, to avoid divergence of $\Delta {\cal S}_u$ ${\cal J}_c$ 
should scale as at least linearly with $T_c$. A stronger version associated to Nernst Heat theorem \cite{nerst06,landsberg89} demands that
${\cal J}_c \propto T_c^{1+\epsilon}$ which ensures the vanishing of entropy production from the cold bath as $T_c \rightarrow 0$ \cite{k278}.
For generic refrigerator models  as $T_c \rightarrow 0$ the cold bath current obtains the universal form:
\begin{equation}
{\cal J}_c  = \hbar {\omega_c}{\cal K} {\cal G}
\end{equation}
where ${\cal K}$ is a heat conductance term. When $T_c \rightarrow 0$ the gain ${\cal G}$ is finite only if $\omega_c \propto T_c$.
This leaves the issue: How does the conductance ${\cal K}$ scale with $\omega_c$, which is model dependent. For example 
for a Bose Einstein Condensate (BEC), the conductance is proportional to the uncondensed fraction leading to ${\cal J}_c \propto T_c^{3}$ \cite{k278,k281}.
\par
The unattainability principle \cite{nerst18},  a different formulation of the III-law states: {\em The zero temperature can be reached 
only if infinite resources are invested}. In quantum mechanics a zero temperature system is in a pure state. This is only possible
if at $T_c \rightarrow 0$,  $\hat \rho_s \otimes \hat \rho_B $. In addition since $\hat \rho_B = |0\rangle \langle 0|$, 
the system-bath interaction energy will vanish, meaning that when the bath approaches its ground state the mechanism of extracting energy
ceases to operate. One can consider two scenarios. Cooling an infinite cold bath and cooling a finite system. Considering an infinite bath
the change in temperature becomes $\frac{d T_c}{dt} = \frac{{ \cal J}_c}{\cal C}$ where ${\cal C}$ is the heat capacity.
This leads to a scaling of $\frac{d T_c}{dt} \propto T_c^{3/2}$ as $T_c \rightarrow 0$ for both degenerate Bose and Fermi gases \cite{k278}.
\par
A different perspective on the III-law can be obtained from quantum resource theory. Using quite general
arguments  the following scaling relation was obtained $\frac{d T_c}{dt} \propto T_c^{1+\frac{1}{7}}$ \cite{masanes2017general}.
Addition of noise can further restrict the minimum achievable temperature \cite{freitas2017fundamental}.
\par
Driven continuous devices require an external power source typically $\hat H_s(t) = \hat H_0 + \hat V f(t)$ where $f(t)$ is a periodic function.
The most direct connection to a driven system is to replace the excitation of the power reservoir by its semiclassical expectation value
$\hat c \sim \bar c e^{-i \omega_w t}$ \cite{k24}. 
A derivation of a thermodynamical consistent  GKLS equation requires to use Floquet theory (\ref{Floq_Dav}) \cite{k122,szczygielski2013markovian}. 
Strong driving alters the system excitations and frequencies Eq. (\ref{FourierS}) which can change the operation conditions from an engine to a dissipator and to a refrigerator \cite{k122,gelbwaser2013minimal}.
\par
Optimizing the performance of continuous devices leads to the tradeoff between efficiency and power. For an engine constructed from
two coupled harmonic oscillators the efficiency at high temperature at maximum power becomes again 
$\eta_{ca}$ Eq. (\ref{ca-efficiency}) \cite{k24}. Universal features of the maximum power efficiency 
have been obtained by information theory considerations \cite{zhou2010minimal}.
\par
Driven systems are prototype models of quantum amplifiers and lasers \cite{gelbwaser2013work}.
Such a treatment ignores the entropy carried away by the amplified light. 
This can be solved by incorporating the emitted light as a single mode harmonic oscillator \cite{boukobza2007three}.
A more complete derivation of the 3-level laser has been derived recently including the entropy dissipated by light\cite{li2017quantum}.

\section{Open Problems and Prospects}

In this section  we discuss  some controversial or unsolved questions of  QT of fundamental nature.

\subsection{Work generation: steady-state or thermodynamic cycles}

One believes that the energy convertors like photovoltaic, thermoelectric and fuel cells, and their biological counterparts can directly transform light, 
heat or chemical energy into work carried by electric current. It is assumed that they do not need any moving parts and operate at non-equilibrium steady-states. 
A recent review of this approach mostly concentrated on the conversion of heat at the nanoscale is given in \cite{benenti2016fundamental}. 
The phenomenological picture of heat to electric current conversion is based on the coupled equations for charge and heat local current densities $j_e$, $j_h$

\begin{eqnarray} 
j_e &=&  \lambda_{ee} \bigl(\nabla \mu /eT\bigr) +  \lambda_{eh} \bigl(\nabla 1/T\bigr)         \nonumber\\
j_e &=&  \lambda_{he} \bigl(\nabla \mu /eT\bigr) +  \lambda_{hh} \bigl(\nabla 1/T\bigr)  
\label{Onsager}
\end{eqnarray}
with local temperature $T$,  local chemical potential $\mu$ and the Onsager matrix $[\lambda_{ab}]$.
\par
The importance of nanoscale devices and the fact that the energy conversion is based on microscopic quantum processes stimulates the development of more fundamental microscopic theories \chqtnt. 
The most popular is stochastic thermodynamics reviewed in \cite{seifert2012stochastic}. Here, the force driving charge carriers is a phenomenological nonconservative force corresponding to a kind of negative friction powered by the external gradients of temperature and chemical potential.
\par
Recently, the steady-state picture has been challenged in a series of papers,  where models of cyclic classical \cite{alicki2017thermodynamic} 
and quantum engines \cite{alicki2015solar,alicki2016thermoelectric} have been proposed. This work was motivated by an apparent inconsistency of \eqref{Onsager} when applied to the devices generating electric current flowing in a closed circuit. Namely, the integral of the steady electric current over its closed path should be different from zero, while the similar integration over the RHS of \eqref{Onsager} yields always zero. 
\par
The "moving parts" in the cyclic models correspond to collective delocalized charge oscillations at the interface of two different materials.
For semiconductor devices they are THz plasma oscillations, while for organic photovoltaic or photosynthetic complexes delocalized and infrared sensitive phonon modes play the role of a ``piston". The cyclic models predict two types of phenomena: emission of coherent radiation by  oscillating "pistons" and reverse effect - enhancement of the generated electric current by coherent resonant radiation stimulating the "piston". In fact, both effects were observed in organic photovoltaic systems \cite{ee2017organic,bakulin2015mode} and the role of selected phonon modes in photosynthesis has also been discussed \cite{chin2013role}.
However, those phenomena were considered as auxiliary effects improving the efficiency of energy converters and not as the necessary elements of their operation principle.

\subsection{Information and thermodynamics}

The idea of representing physical processes as computation processes or more generally as information processing is quite popular  in the rapidly developing field of quantum information \cite{nielsen2002quantum}. For example, one believes that an acquired bit of information can be traded-off for a $k_B T \ln 2$ of work extracted from the bath at temperature $T$. 
This leads to the Landauer formula \cite{Landauer:1961,Bennett:2003}
which puts the lower limit equal to $k_B T \ln 2$ for the work needed to reset a bit of information in a memory \chqfluct. The reasoning is based on the idea of Szilard \cite{szilard1925,szilard1929} who proposed a model of an engine which consists of a box with a single gas particle, in thermal contact with a heat bath, and a partition. The partition can be inserted into the box, dividing it into two equal volumes, and can slide without friction along the box. To extract  $k_B T\ln 2$ of work in an isothermal process of gas expansion one connects up the partition to a pulley. Szilard assumed that in order  to realize work extraction it is necessary to know  "which side the molecule is on'' which corresponds to one bit of information. This model was generalized in various directions including the quantum case, \cite{Sagawa:2012} and claimed to be realized experimentally in the classical \cite{Toyabe}, \cite{Lutz} and the quantum domain \cite{pekola2015towards}.
\par
However, the very idea of the equivalence between information and thermodynamics remains controversial \cite{Norton}, \cite{Shenker} \cite{alicki2014information}. As noticed already by Popper and Feyerabend \cite{Feyerabend} there exist procedures of extracting work without knowing the position of the particle and, on the other hand the mechanism of inserting  a partition can provide  a necessary amount of work to avoid the conflict with the Kelvin formulation of the Second Law.
\par
The recently developed resource theory of quantum thermodynamics is another
example of the interplay between information theory and thermodynamics \cite{horodecki2013fundamental,gour2015resource,goold2016role,vinjanampathy2016quantum} \chrto~ \chthinf.
The theory is an axiomatic approach with a mathematical structure motivated by the theory of entanglement.
The resource in this theory are states with informational nonequillibrium. 
Resource theories in quantum information identify a set of restrictive operations that can act on Òvaluable resource
statesÓ. For a given initial state these restrictive operations then define a set of states that are
reachable. For example energy conserving  unitaries on the system bath and work repository.
The single shot regime refers to operating on a single quantum system, which can be a highly
correlated system of many subsystems, rather than on an infinite ensemble of identical and independently
distributed copies of a quantum system
\cite{vinjanampathy2016quantum,korzekwa2016extraction,goold16,brandao2015second}.
The idea is to find additional restrictions on possible thermodynamical transformation on finite systems.
For example single shot II-laws based on properties of  R{\'e}nyi divergence \cite{renyi1961measures,brandao2015second} 
which in the thermodynamical limit  converge to the standard II-law.
The drawback of the theory is that there is no dynamics so there is no reference to a fast or slow operation.

.

\subsection{Work and Heat}

One of the great discoveries in the history of science was the recognition that heat is a form of energy. This allowed to interpret the phenomenological First Law of Thermodynamics 
\begin{equation}
dE = \delta Q - \delta W 
\label{ILaw}
\end{equation}
as the instance of \emph{energy conservation principle}. However, in contrast to internal energy $E$, indentified 
with the total energy of the system, work $W$ and heat $Q$
are path-dependent and are therefore  \emph{thermodynamic process functions}. In quantum language 
it means that there are neither described by hermitian operators \cite{talkner2007fluctuation} 
nor by nonlinear functions of density matrices like e.g. von Neumann entropy. 
\par
It seems that, generally, the instantaneous decomposition corresponding to \eqref{ILaw} 
may be even impossible as one needs certain time-scale to decide which part of energy is related to a random motion (heat) 
or to a deterministic one (work). It seems also that heat, 
which is transported by irreversible  processes can be determined  easier than work. 
The case of Markovian dynamics illustrates very well this problem. Only for slow
driving there exists a  natural "instantaneous'' analog of \eqref{ILaw} given by \eqref{work_heat}, 
while for fast periodic driving only temporal heat currents are well-defined and the unique form of the I-Law is known in the limit cycle only \eqref{power_per}.
\par
There exists a  number of proposals in the literature to define work and heat beyond the Markovian approximation:
\begin{enumerate}
\item{Work defined in terms of two measurements \cite{roncaglia2014work,sampaio2017impossible} 
(useful for Hamiltonian dynamics, fluctuation theorems and full counting statistics \chflqwork \chqtraj).}
\item{
Heat as the energy exchanged with a bath (including the assumptions of good ergodic properties of the bath and weak influence on the bath by the system).}
\item{
Work reservoir represented by a sink (e.g. low lying level) in Markovian Master equations.
Here, the applied standard entropy balance suggests that the energy flow  $J_{sink}$  
dumped in a sink is accompanied by a large flow of entropy  $J_{en} =  J_{sink}/T_{sink}$  with the effective sink temperature usually close or even equal to zero. 
It suggests that $J_{sink} $ should be rather interpreted as a heat flow, as work is "energy with negligible entropy". 
Otherwise  violation of the second law can occur  \cite{gelbwaser2017thermodynamic}.}
\item{
Work reservoir represented by a quantum system (e.g. harmonic oscillator). 
Treating the whole transferred energy as work one can violate the Carnot bound \cite{boukobza2013breaking}.
The proper procedure seems to be using ergotropy as a measure of work stored in the work reservoir \cite{alicki2017gkls} \chtpi \chqbat.}
\item{
Work measured by \emph{wits} (qubits in excited state), resource theory, analogy to qubits, \cite{horodecki2013fundamental}
.}
\end{enumerate}

\subsection{Thermalization}

Considering a finite quantum system: What are the properties that it can serve as a bath? How large does it have to be?
what should be its spectrum? how should it couple to the system? Does the system bath dynamics mimic the Markovian GKLS dynamics?
\par
Thermalization can be described as a process where the system loses its memory partly or completely of its initial state and the system settles to
a steady state. In classical mechanics chaotic dynamics even in a finite system are sufficient to lead to thermalization.
On the contrary, an isolated quantum system has a discrete spectrum and therefore its dynamics is quasiperiodic. Thus 
strictly speaking, in terms of positive Kolmogorov entropy isolated quantum systems are non chaotic \cite{k11}. 
Another property should lead to quantum thermalization.
\par
The eigenvalue thermalization hypothesis (ETH) \cite{deutsch1991quantum,srednicki1994chaos} \chdtyp~\chnmbq, 
applies for strongly coupled quantum systems which therefore possess a Wigner-Dyson
distribution of energy gaps \cite{dyson1962statistical}. The conjecture is that the expectation value of any operator 
$\hat A$ will relax asymptotically to its microcanonical value, with the notation of Eq. (\ref{ergodic_av}) 
and $A_{jj}=\langle j | \hat A| j \rangle$\cite{rigol2008thermalization}:
\begin{eqnarray}
\sum_{j} |c_{j}|^2 A_{j j} = \langle A \rangle_{microcan} (E_0)= \frac{1}{{\cal N}_{E_0 \Delta E}} 
\sum_{\substack{{j}\\{|E_0-E_{j}|< \Delta E}} }A_{j j}
\label{eq:ETH}
\end{eqnarray}
where $E_0$ is the mean energy of the initial state, $\Delta E$ is the half-width
of an appropriately chosen energy window centred at $E_0$, and ${\cal N}_{E_0 \Delta E}$ the normalization. 
The ETH hypothesis has been extensively tested numerically
and has been found to apply in sufficiently large and complex systems
\cite{rigol2008thermalization,kim2014testing,steinigeweg2014pushing,ikeda2013finite,alba2015eigenstate}.
One should comment that the popular bath composed on noninteracting harmonic oscillators does not
fulfil the requirements of the eigenvalue thermalization hypothesis, its ergodic properties being weak
because it is a quasi-free system with additional constants of motion.
\par
We can now apply the ETH to a small quantum system coupled to a finite strongly coupled bath. 
In this case we expect the system to converge to a canonical state.
The operators of interest are local in the system.
Therefore according to the ETH we expect them to relax to a value which is determined by the bath mean energy 
with a correction to the finite heat capacity of the bath. This idea has been tested for a system consisting of a one 
and two qubits and a bath consisting of 32 or 34 strongly and randomly coupled spins. 
The initial state of the bath was a random phase thermal wavefunction. 
A Hilbert size of $\sim 10^{11}$ employed for the study is on the limit of simulation by currently available classical computers.
The ETH proved to be correct with respect to the asymptotic system expectation values \cite{zhao2016dynamics,de2017relaxation}.
In addition, for the one qubit case a Bloch-type equation with time-dependent coefficients provides a simple and accurate description 
of the dynamics of a spin particle in contact with a thermal bath. A similar result was found for the 2-qubit system with a variety of bath models.

\subsection{Concluding remarks}

The  recent rapid development in the  field of quantum thermodynamics is  intimately connected to the quantum theory of open systems 
and strongly influenced by the ideas and methods of quantum information. The new directions of theoretical research are stimulated 
by the fast technological progress in construction and precise control of micro(meso)scopic devices for information 
processing and energy transduction. The  implementations  cover a vast spectrum of physical systems including quantum optical, 
superconducting, solid state or based on organic molecules devices.  The operation conditions for all these systems requires refrigeration.
\par
These  emerging technologies  pose  problems of  reliability, scalability   and efficiency,  
related to the fundamental principles of thermodynamics, which have to be properly extended to the quantum domain. 
This extension is a highly nontrivial and controversial task because the standard simplifications used for macroscopic 
systems are generally not valid at  micro(meso) scopic scale,  short time-scales and  at the presence of strong correlations.  
Therefore, even the unique definitions of fundamental notions like heat, work and entropy are available only in the limiting cases. 
Although most of the results suggest that the laws of thermodynamics are still valid in the averaged sense, the role of quantum effects remains an open problem.
\par
One can expect that the further analysis of particular models  like quantum heat/chemical engines, quantum pumps, 
quantum clocks or  quantum switches, including mechanisms of  feedback and self-oscillations should 
provide new inputs for improvement and new designs of quantum thermodynamic machines.

\section{Acknowledgments}
We want to thank Amikam Levi and Raam Uzdin for their helpful comments. The work was partially supported by the Israeli Science Foundation: Grant 2244/14. 
\break
\bibliography{dephc1,pub}

\begin{thebibliography}{100}

\bibitem{planck1900}
Max Karl Ernst~Ludwig Planck.
\newblock Zur theorie des gesetzes der energieverteilung im normalspectrum.
\newblock {\em Verhandl. Dtsc. Phys. Ges.}, 2:237, 1900.

\bibitem{einstein05}
Albert Einstein.
\newblock {\"U}ber einen die erzeugung und verwandlung des lichtes betreffenden
  heuristischen gesichtspunkt.
\newblock {\em Annalen der physik}, 322(6):132--148, 1905.

\bibitem{vNeumann}
John~Von Neumann.
\newblock {\em Mathematical foundations of quantum mechanics}.
\newblock Number~2. Princeton university press, 1955.

\bibitem{einstein1916}
Albert Einstein.
\newblock Strahlungs-emission und absorption nach der quantentheorie.
\newblock {\em Deutsche Physikalische Gesellschaft}, 18, 1916.

\bibitem{scovil1959}
HED Scovil and EO~Schulz-DuBois.
\newblock Three-level masers as heat engines.
\newblock {\em Physical Review Letters}, 2(6):262, 1959.

\bibitem{geusic1959}
JE~Geusic, BO~Schulz-DuBois, RW~De~Grasse, and HED Scovil.
\newblock Three level spin refrigeration and maser action at 1500 mc/sec.
\newblock {\em Journal of Applied Physics}, 30(7):1113--1114, 1959.

\bibitem{geusic1967quantum}
JE~Geusic, EO~Schulz-DuBios, and HED Scovil.
\newblock Quantum equivalent of the carnot cycle.
\newblock {\em Physical Review}, 156(2):343, 1967.

\bibitem{wineland1978}
David~J Wineland, Robert~E Drullinger, and Fred~L Walls.
\newblock Radiation-pressure cooling of bound resonant absorbers.
\newblock {\em Physical Review Letters}, 40(25):1639, 1978.

\bibitem{hansch1975}
Theodor~W H{\"a}nsch and Arthur~L Schawlow.
\newblock Cooling of gases by laser radiation.
\newblock {\em Optics Communications}, 13(1):68--69, 1975.

\bibitem{heisenberg1949physical}
Werner Heisenberg.
\newblock {\em The physical principles of the quantum theory}.
\newblock Courier Corporation, 1949.

\bibitem{schrodinger1926undulatory}
Erwin Schr{\"o}dinger.
\newblock An undulatory theory of the mechanics of atoms and molecules.
\newblock {\em Physical Review}, 28(6):1049, 1926.

\bibitem{dirac1981principles}
Paul Adrien~Maurice Dirac.
\newblock {\em The principles of quantum mechanics}.
\newblock Number~27. Oxford university press, 1981.

\bibitem{rigol2008thermalization}
Marcos Rigol, Vanja Dunjko, and Maxim Olshanii.
\newblock Thermalization and its mechanism for generic isolated quantum
  systems.
\newblock {\em Nature}, 452(7189):854--858, 2008.

\bibitem{callen1998}
Herbert~B Callen.
\newblock Thermodynamics and an introduction to thermostatistics, 1998.

\bibitem{haag1974}
Rudolf Haag, Daniel Kastler, and Ewa~B Trych-Pohlmeyer.
\newblock Stability and equilibrium states.
\newblock {\em Communications in Mathematical Physics}, 38(3):173--193, 1974.

\bibitem{lenard1978}
A~Lenard.
\newblock Thermodynamical proof of the gibbs formula for elementary quantum
  systems.
\newblock {\em Journal of Statistical Physics}, 19(6):575--586, 1978.

\bibitem{pusz1978passive}
W~Pusz and SL~Woronowicz.
\newblock Passive states and kms states for general quantum systems.
\newblock {\em Communications in Mathematical Physics}, 58(3):273--290, 1978.

\bibitem{kubo1957}
Ryogo Kubo.
\newblock Statistical-mechanical theory of irreversible processes. i. general
  theory and simple applications to magnetic and conduction problems.
\newblock {\em Journal of the Physical Society of Japan}, 12(6):570--586, 1957.

\bibitem{green1954markoff}
Melville~S Green.
\newblock Markoff random processes and the statistical mechanics of
  time-dependent phenomena. ii. irreversible processes in fluids.
\newblock {\em The Journal of Chemical Physics}, 22(3):398--413, 1954.

\bibitem{martin1959}
Paul~C Martin and Julian Schwinger.
\newblock Theory of many-particle systems. i.
\newblock {\em Physical Review}, 115(6):1342, 1959.

\bibitem{bratteli1996operator}
Ola Bratteli and Derek~W Robinson.
\newblock Operator algebras and quantum statistical mechanics. vol. 2:
  Equilibrium states. models in quantum statistical mechanics.
\newblock 1996.

\bibitem{Kraus71}
Karl Kraus.
\newblock General state changes in quantum theory.
\newblock {\em Annals of Physics}, 64(2):311--335, 1971.

\bibitem{stinespring1955}
W~Forrest Stinespring.
\newblock Mr0069403 (16, 1033b) 46.0 x.
\newblock In {\em Proc. Amer. Math. Soc}, volume~6, pages 211--216, 1955.

\bibitem{lindblad75}
G.~Lindblad.
\newblock Completely positive maps and entropy inequalities.
\newblock {\em Comm. Math. Phys.}, 40:147, 1975.

\bibitem{chruscinski2017brief}
Dariusz Chru{\'s}ci{\'n}ski and Saverio Pascazio.
\newblock A brief history of the gkls equation.
\newblock {\em Open Systems \& Information Dynamics}, 24(03):1740001, 2017.

\bibitem{gorini1976completely}
Vittorio Gorini, Andrzej Kossakowski, and Ennackal Chandy~George Sudarshan.
\newblock Completely positive dynamical semigroups of n-level systems.
\newblock {\em Journal of Mathematical Physics}, 17(5):821--825, 1976.

\bibitem{lindblad76}
Goran Lindblad.
\newblock On the generators of quantum dynamical semigroups.
\newblock {\em Communications in Mathematical Physics}, 48(2):119--130, 1976.

\bibitem{siemon2017unbounded}
Inken Siemon, Alexander~S Holevo, and Reinhard~F Werner.
\newblock Unbounded generators of dynamical semigroups.
\newblock {\em Open Systems \& Information Dynamics}, 24(04):1740015, 2017.

\bibitem{davies74}
E~Brian Davies.
\newblock Markovian master equations.
\newblock {\em Communications in mathematical Physics}, 39(2):91--110, 1974.

\bibitem{wangsness1953dynamical}
Roald~K Wangsness and Felix Bloch.
\newblock The dynamical theory of nuclear induction.
\newblock {\em Physical Review}, 89(4):728, 1953.

\bibitem{redfield1957theory}
Alfred~G Redfield.
\newblock On the theory of relaxation processes.
\newblock {\em IBM Journal of Research and Development}, 1(1):19--31, 1957.

\bibitem{nakajima1958quantum}
Sadao Nakajima.
\newblock On quantum theory of transport phenomena: steady diffusion.
\newblock {\em Progress of Theoretical Physics}, 20(6):948--959, 1958.

\bibitem{zwanzig1960ensemble}
Robert Zwanzig.
\newblock Ensemble method in the theory of irreversibility.
\newblock {\em The Journal of Chemical Physics}, 33(5):1338--1341, 1960.

\bibitem{fermi1950nuclear}
Enrico Fermi.
\newblock {\em Nuclear physics: a course given by Enrico Fermi at the
  University of Chicago}.
\newblock University of Chicago Press, 1950.

\bibitem{alicki1977markov}
Robert Alicki.
\newblock The markov master equations and the fermi golden rule.
\newblock {\em International Journal of Theoretical Physics}, 16(5):351--355,
  1977.

\bibitem{frigerio1977quantum}
Alberto Frigerio.
\newblock Quantum dynamical semigroups and approach to equilibrium.
\newblock {\em Letters in Mathematical Physics}, 2(2):79--87, 1977.

\bibitem{davies1978}
EB~Davies and H~Spohn.
\newblock Open quantum systems with time-dependent hamiltonians and their
  linear response.
\newblock {\em Journal of Statistical Physics}, 19(5):511--523, 1978.

\bibitem{alicki1979quantum}
Robert Alicki.
\newblock The quantum open system as a model of the heat engine.
\newblock {\em Journal of Physics A: Mathematical and General}, 12(5):L103,
  1979.

\bibitem{spohn1978entropy}
Herbert Spohn.
\newblock Entropy production for quantum dynamical semigroups.
\newblock {\em Journal of Mathematical Physics}, 19(5):1227--1230, 1978.

\bibitem{mcadory1977}
RT~McAdory~Jr and WC~Schieve.
\newblock On entropy production in a stochastic model of open systems.
\newblock {\em The Journal of Chemical Physics}, 67(5):1899--1903, 1977.

\bibitem{spohn1978irreversible}
Herbert Spohn and Joel~L Lebowitz.
\newblock Irreversible thermodynamics for quantum systems weakly coupled to
  thermal reservoirs.
\newblock {\em Adv. Chem. Phys}, 38:109--142, 1978.

\bibitem{k114}
Eitan Geva, Ronnie Kosloff, and JL~Skinner.
\newblock On the relaxation of a two-level system driven by a strong
  electromagnetic field.
\newblock {\em The Journal of Chemical Physics}, 102(21):8541--8561, 1995.

\bibitem{kohler1997floquet}
Sigmund Kohler, Thomas Dittrich, and Peter H{\"a}nggi.
\newblock Floquet-markovian description of the parametrically driven,
  dissipative harmonic quantum oscillator.
\newblock {\em Physical Review E}, 55(1):300, 1997.

\bibitem{k275}
Amikam Levy, Robert Alicki, and Ronnie Kosloff.
\newblock Quantum refrigerators and the third law of thermodynamics.
\newblock {\em Physical Review E}, 85:061126, 2012.

\bibitem{szczygielski2013markovian}
Krzysztof Szczygielski, David Gelbwaser-Klimovsky, and Robert Alicki.
\newblock Markovian master equation and thermodynamics of a two-level system in
  a strong laser field.
\newblock {\em Physical Review E}, 87(1):012120, 2013.

\bibitem{alicki2015non}
Robert Alicki and David Gelbwaser-Klimovsky.
\newblock Non-equilibrium quantum heat machines.
\newblock {\em New Journal of Physics}, 17(11):115012, 2015.

\bibitem{rossnagel2014nanoscale}
Johannes Ro{\ss}nagel, Obinna Abah, Ferdinand Schmidt-Kaler, Kilian Singer, and
  Eric Lutz.
\newblock Nanoscale heat engine beyond the carnot limit.
\newblock {\em Physical review letters}, 112(3):030602, 2014.

\bibitem{niedenzu2017universal}
Wolfgang Niedenzu, Victor Mukherjee, Arnab Ghosh, Abraham~G Kofman, and Gershon
  Kurizki.
\newblock Universal thermodynamic limit of quantum engine efficiency.
\newblock {\em arXiv preprint arXiv:1703.02911}, 2017.

\bibitem{k281}
Ronnie Kosloff.
\newblock Quantum thermodynamics: a dynamical viewpoint.
\newblock {\em Entropy}, 15(6):2100--2128, 2013.

\bibitem{pechukas1994reduced}
Philip Pechukas.
\newblock Reduced dynamics need not be completely positive.
\newblock {\em Physical review letters}, 73(8):1060, 1994.

\bibitem{alicki1995comment}
Robert Alicki.
\newblock Comment on �reduced dynamics need not be completely positive�.
\newblock {\em Physical review letters}, 75(16):3020, 1995.

\bibitem{lindblad1996existence}
G{\"o}ran Lindblad.
\newblock On the existence of quantum subdynamics.
\newblock {\em Journal of Physics A: Mathematical and General}, 29(14):4197,
  1996.

\bibitem{argentieri2014violations}
Giuseppe Argentieri, Fabio Benatti, Roberto Floreanini, and Marco Pezzutto.
\newblock Violations of the second law of thermodynamics by a non-completely
  positive dynamics.
\newblock {\em EPL (Europhysics Letters)}, 107(5):50007, 2014.

\bibitem{argentieri2015complete}
Giuseppe Argentieri, Fabio Benatti, Roberto Floreanini, and Marco Pezzutto.
\newblock Complete positivity and thermodynamics in a driven open quantum
  system.
\newblock {\em Journal of Statistical Physics}, 159(5):1127--1153, 2015.

\bibitem{caldeira1983path}
Amir~O Caldeira and Anthony~J Leggett.
\newblock Path integral approach to quantum brownian motion.
\newblock {\em Physica A: Statistical mechanics and its Applications},
  121(3):587--616, 1983.

\bibitem{diosi1993calderia}
Lajos Di{\'o}si.
\newblock Calderia-leggett master equation and medium temperatures.
\newblock {\em Physica A: Statistical Mechanics and its Applications},
  199(3-4):517--526, 1993.

\bibitem{kohen1997phase}
Daniela Kohen, C~Clay Marston, and David~J Tannor.
\newblock Phase space approach to theories of quantum dissipation.
\newblock {\em The Journal of chemical physics}, 107(13):5236--5253, 1997.

\bibitem{lindblad1976brownian}
G{\"o}ran Lindblad.
\newblock Brownian motion of a quantum harmonic oscillator.
\newblock {\em Reports on Mathematical Physics}, 10(3):393--406, 1976.

\bibitem{k290}
Amikam Levy and Ronnie Kosloff.
\newblock The local approach to quantum transport may violate the second law of
  thermodynamics.
\newblock {\em EPL (Europhysics Letters)}, 107(2):20004, 2014.

\bibitem{hofer2017markovian}
Patrick~P Hofer, Mart{\'\i} Perarnau-Llobet, L~David~M Miranda, G{\'e}raldine
  Haack, Ralph Silva, Jonatan~Bohr Brask, and Nicolas Brunner.
\newblock Markovian master equations for quantum thermal machines: local vs
  global approach.
\newblock {\em New Journal of Physics}, 19:123037, 2017.

\bibitem{mitchison2017non}
Mark~Thomas Mitchison and Martin~Bodo Plenio.
\newblock Non-additive dissipation in open quantum networks out of equilibrium.
\newblock {\em New Journal of Physics}, 20:033005, 2018.

\bibitem{gonzalez2017testing}
J~Onam Gonz{\'a}lez, Luis~A Correa, Giorgio Nocerino, Jos{\'e}~P Palao, Daniel
  Alonso, and Gerardo Adesso.
\newblock Testing the validity of the �local�and �global�gkls master
  equations on an exactly solvable model.
\newblock {\em Open Systems \& Information Dynamics}, 24(04):1740010, 2017.

\bibitem{brask2017additivity}
Jonatan~Bohr Brask, Jan Ko{\l}ody{\'n}ski, Mart{\'\i} Perarnau-Llobet, and
  Bogna Bylicka.
\newblock Additivity of dynamical generators for quantum master equations.
\newblock {\em arXiv preprint arXiv:1704.08702}, 2017.

\bibitem{k122}
Eitan Geva and Ronnie Kosloff.
\newblock The quantum heat engine and heat pump: An irreversible thermodynamic
  analysis of the three-level amplifier.
\newblock {\em The Journal of Chemical Physics}, 104(19):7681--7699, 1996.

\bibitem{haake1973statistical}
Fritz Haake.
\newblock Statistical treatment of open systems by generalized master
  equations.
\newblock In {\em Springer tracts in modern physics}, pages 98--168. Springer,
  1973.

\bibitem{kleinekathofer2004non}
Ulrich Kleinekath{\"o}fer.
\newblock Non-markovian theories based on a decomposition of the spectral
  density.
\newblock {\em The Journal of chemical physics}, 121(6):2505--2514, 2004.

\bibitem{de2017dynamics}
In{\'e}s de~Vega and Daniel Alonso.
\newblock Dynamics of non-markovian open quantum systems.
\newblock {\em Reviews of Modern Physics}, 89(1):015001, 2017.

\bibitem{shibata1977generalized}
Fumiaki Shibata, Yoshinori Takahashi, and Natsuki Hashitsume.
\newblock A generalized stochastic liouville equation. non-markovian versus
  memoryless master equations.
\newblock {\em Journal of Statistical Physics}, 17(4):171--187, 1977.

\bibitem{kofman2004unified}
AG~Kofman and G~Kurizki.
\newblock Unified theory of dynamically suppressed qubit decoherence in thermal
  baths.
\newblock {\em Physical review letters}, 93(13):130406, 2004.

\bibitem{tanimura1989time}
Yoshitaka Tanimura and Ryogo Kubo.
\newblock Time evolution of a quantum system in contact with a nearly
  gaussian-markoffian noise bath.
\newblock {\em Journal of the Physical Society of Japan}, 58(1):101--114, 1989.

\bibitem{meier1999non}
Christoph Meier and David~J Tannor.
\newblock Non-markovian evolution of the density operator in the presence of
  strong laser fields.
\newblock {\em The Journal of chemical physics}, 111(8):3365--3376, 1999.

\bibitem{jin2008exact}
Jinshuang Jin, Xiao Zheng, and YiJing Yan.
\newblock Exact dynamics of dissipative electronic systems and quantum
  transport: Hierarchical equations of motion approach.
\newblock {\em The Journal of chemical physics}, 128(23):234703, 2008.

\bibitem{ishizaki2009unified}
Akihito Ishizaki and Graham~R Fleming.
\newblock Unified treatment of quantum coherent and incoherent hopping dynamics
  in electronic energy transfer: Reduced hierarchy equation approach.
\newblock {\em The Journal of chemical physics}, 130(23):234111, 2009.

\bibitem{marcantoni2017entropy}
STEFANO Marcantoni, S~Alipour, FABIO Benatti, R~Floreanini, and AT~Rezakhani.
\newblock Entropy production and non-markovian dynamical maps.
\newblock {\em Scientific reports}, 7(1):12447, 2017.

\bibitem{alipour2016correlations}
S~Alipour, F~Benatti, F~Bakhshinezhad, M~Afsary, S~Marcantoni, and
  AT~Rezakhani.
\newblock Correlations in quantum thermodynamics: Heat, work, and entropy
  production.
\newblock {\em Scientific reports}, 6, 2016.

\bibitem{esposito2015quantum}
Massimiliano Esposito, Maicol~A Ochoa, and Michael Galperin.
\newblock Quantum thermodynamics: A nonequilibrium green?s function approach.
\newblock {\em Physical review letters}, 114(8):080602, 2015.

\bibitem{ludovico2016dynamics}
Mar{\'\i}a~Florencia Ludovico, Michael Moskalets, David S{\'a}nchez, and
  Liliana Arrachea.
\newblock Dynamics of energy transport and entropy production in ac-driven
  quantum electron systems.
\newblock {\em Physical Review B}, 94(3):035436, 2016.

\bibitem{ochoa2016energy}
Maicol~A Ochoa, Anton Bruch, and Abraham Nitzan.
\newblock Energy distribution and local fluctuations in strongly coupled open
  quantum systems: The extended resonant level model.
\newblock {\em Physical Review B}, 94(3):035420, 2016.

\bibitem{schaller2013single}
Gernot Schaller, Thilo Krause, Tobias Brandes, and Massimiliano Esposito.
\newblock Single-electron transistor strongly coupled to vibrations: counting
  statistics and fluctuation theorem.
\newblock {\em New Journal of Physics}, 15(3):033032, 2013.

\bibitem{segal2014two}
Dvira Segal.
\newblock Two-level system in spin baths: Non-adiabatic dynamics and heat
  transport.
\newblock {\em The Journal of chemical physics}, 140(16):164110, 2014.

\bibitem{xu2016polaron}
Dazhi Xu, Chen Wang, Yang Zhao, and Jianshu Cao.
\newblock Polaron effects on the performance of light-harvesting systems: a
  quantum heat engine perspective.
\newblock {\em New Journal of Physics}, 18(2):023003, 2016.

\bibitem{wang2017unifying}
Chen Wang, Jie Ren, and Jianshu Cao.
\newblock Unifying quantum heat transfer in a nonequilibrium spin-boson model
  with full counting statistics.
\newblock {\em Physical Review A}, 95(2):023610, 2017.

\bibitem{gelbwaser2015strongly}
David Gelbwaser-Klimovsky and Al{\'a}n Aspuru-Guzik.
\newblock Strongly coupled quantum heat machines.
\newblock {\em The journal of physical chemistry letters}, 6(17):3477--3482,
  2015.

\bibitem{bruch2016quantum}
Anton Bruch, Mark Thomas, Silvia~Viola Kusminskiy, Felix von Oppen, and Abraham
  Nitzan.
\newblock Quantum thermodynamics of the driven resonant level model.
\newblock {\em Physical Review B}, 93(11):115318, 2016.

\bibitem{strasberg2016nonequilibrium}
Philipp Strasberg, Gernot Schaller, Neill Lambert, and Tobias Brandes.
\newblock Nonequilibrium thermodynamics in the strong coupling and
  non-markovian regime based on a reaction coordinate mapping.
\newblock {\em New Journal of Physics}, 18(7):073007, 2016.

\bibitem{newman2017performance}
David Newman, Florian Mintert, and Ahsan Nazir.
\newblock Performance of a quantum heat engine at strong reservoir coupling.
\newblock {\em Physical Review E}, 95(3):032139, 2017.

\bibitem{perarnau2018strong}
M~Perarnau-Llobet, H~Wilming, A~Riera, R~Gallego, and J~Eisert.
\newblock Strong coupling corrections in quantum thermodynamics.
\newblock {\em Physical Review Letters}, 120(12):120602, 2018.

\bibitem{k307}
Gil Katz and Ronnie Kosloff.
\newblock Quantum thermodynamics in strong coupling: Heat transport and
  refrigeration.
\newblock {\em Entropy}, 18(5):186, 2016.

\bibitem{k306}
Raam Uzdin, Amikam Levy, and Ronnie Kosloff.
\newblock Quantum heat machines equivalence and work extraction beyond
  markovianity, and strong coupling via heat exchangers.
\newblock {\em Entropy}, 18:124, 2016.

\bibitem{carnot1872}
Sadi Carnot.
\newblock R{\'e}flexions sur la puissance motrice du feu et sur les machines
  propres {\`a} d{\'e}velopper cette puissance.
\newblock Annales scientifiques de l'Ecole normale, 1872.

\bibitem{Rossnagel325}
Johannes Ro{\ss}nagel, Samuel~T. Dawkins, Karl~N. Tolazzi, Obinna Abah, Eric
  Lutz, Ferdinand Schmidt-Kaler, and Kilian Singer.
\newblock A single-atom heat engine.
\newblock {\em Science}, 352(6283):325--329, 2016.

\bibitem{k201}
Tova Feldmann and Ronnie Kosloff.
\newblock Characteristics of the limit cycle of a reciprocating quantum heat
  engine.
\newblock {\em Physical Review E}, 70(4):046110, 2004.

\bibitem{k221}
Yair Rezek and Ronnie Kosloff.
\newblock Irreversible performance of a quantum harmonic heat engine.
\newblock {\em New Journal of Physics}, 8(5):83, 2006.

\bibitem{k85}
Eitan Geva and Ronnie Kosloff.
\newblock A quantum-mechanical heat engine operating in finite time. a model
  consisting of spin-1/2 systems as the working fluid.
\newblock {\em The Journal of Chemical Physics}, 96(4):3054--3067, 1992.

\bibitem{k190}
Tova Feldmann and Ronnie Kosloff.
\newblock Quantum four-stroke heat engine: Thermodynamic observables in a model
  with intrinsic friction.
\newblock {\em Physical Review E}, 68(1):016101, 2003.

\bibitem{esposito2012stochastic}
Massimiliano Esposito.
\newblock Stochastic thermodynamics under coarse graining.
\newblock {\em Physical Review E}, 85(4):041125, 2012.

\bibitem{seifert2012stochastic}
Udo Seifert.
\newblock Stochastic thermodynamics, fluctuation theorems and molecular
  machines.
\newblock {\em Reports on Progress in Physics}, 75(12):126001, 2012.

\bibitem{plastina2014}
F~Plastina, A~Alecce, TJG Apollaro, G~Falcone, G~Francica, F~Galve, N~Lo Gullo,
  and R~Zambrini.
\newblock Irreversible work and inner friction in quantum thermodynamic
  processes.
\newblock {\em Physical Review Letters}, 113(26):260601, 2014.

\bibitem{muga09}
Xi~Chen, A~Ruschhaupt, S~Schmidt, A~Del~Campo, David Gu{\'e}ry-Odelin, and
  J~Gonzalo Muga.
\newblock Fast optimal frictionless atom cooling in harmonic traps: Shortcut to
  adiabaticity.
\newblock {\em Physical review letters}, 104(6):063002, 2010.

\bibitem{torrontegui2013shortcuts}
Erik Torrontegui, Sara Ib{\'a}nez, Sofia Mart{\'\i}nez-Garaot, Michele Modugno,
  Adolfo del Campo, David Gu{\'e}ry-Odelin, Andreas Ruschhaupt, Xi~Chen,
  Juan~Gonzalo Muga, et~al.
\newblock Shortcuts to adiabaticity.
\newblock {\em Adv. At. Mol. Opt. Phys}, 62:117--169, 2013.

\bibitem{del2014more}
Adolfo Del~Campo, J~Goold, and M~Paternostro.
\newblock More bang for your buck: Super-adiabatic quantum engines.
\newblock {\em Scientific reports}, 4, 2014.

\bibitem{anandan1990geometry}
J~Anandan and Yakir Aharonov.
\newblock Geometry of quantum evolution.
\newblock {\em Physical review letters}, 65(14):1697, 1990.

\bibitem{deffner2013energy}
Sebastian Deffner and Eric Lutz.
\newblock Energy--time uncertainty relation for driven quantum systems.
\newblock {\em Journal of Physics A: Mathematical and Theoretical},
  46(33):335302, 2013.

\bibitem{chen2010fast}
Xi~Chen, A~Ruschhaupt, S~Schmidt, A~Del~Campo, David Gu{\'e}ry-Odelin, and
  J~Gonzalo Muga.
\newblock Fast optimal frictionless atom cooling in harmonic traps: Shortcut to
  adiabaticity.
\newblock {\em Physical review letters}, 104(6):063002, 2010.

\bibitem{stefanatos2017minimum}
Dionisis Stefanatos.
\newblock Minimum-time transitions between thermal and fixed average energy
  states of the quantum parametric oscillator.
\newblock {\em SIAM Journal on Control and Optimization}, 55(3):1429--1451,
  2017.

\bibitem{scully2003extracting}
Marlan~O Scully, M~Suhail Zubairy, Girish~S Agarwal, and Herbert Walther.
\newblock Extracting work from a single heat bath via vanishing quantum
  coherence.
\newblock {\em Science}, 299(5608):862--864, 2003.

\bibitem{niedenzu2016operation}
Wolfgang Niedenzu, David Gelbwaser-Klimovsky, Abraham~G Kofman, and Gershon
  Kurizki.
\newblock On the operation of machines powered by quantum non-thermal baths.
\newblock {\em New Journal of Physics}, 18(8):083012, 2016.

\bibitem{novikov1958efficiency}
II~Novikov.
\newblock The efficiency of atomic power stations (a review).
\newblock {\em Journal of Nuclear Energy (1954)}, 7(1):125--128, 1954.

\bibitem{curzon75}
FL~Curzon and B~Ahlborn.
\newblock Efficiency of a carnot engine at maximum power output.
\newblock {\em American Journal of Physics}, 43(1):22--24, 1975.

\bibitem{k294}
Raam Uzdin and Ronnie Kosloff.
\newblock Universal features in the efficiency at maximal work of hot quantum
  otto engines.
\newblock {\em EPL (Europhysics Letters)}, 108(4):40001, 2014.

\bibitem{cavina2017slow}
Vasco Cavina, Andrea Mari, and Vittorio Giovannetti.
\newblock Slow dynamics and thermodynamics of open quantum systems.
\newblock {\em Physical review letters}, 119(5):050601, 2017.

\bibitem{andresen1977}
Bjarne Andresen, R~Stephen Berry, Abraham Nitzan, and Peter Salamon.
\newblock Thermodynamics in finite time. i. the step-carnot cycle.
\newblock {\em Physical Review A}, 15(5):2086, 1977.

\bibitem{salamon01}
Peter Salamon, JD~Nulton, Gino Siragusa, Torben~R Andersen, and Alfonso Limon.
\newblock Principles of control thermodynamics.
\newblock {\em Energy}, 26(3):307--319, 2001.

\bibitem{k87}
Eitan Geva and Ronnie Kosloff.
\newblock On the classical limit of quantum thermodynamics in finite time.
\newblock {\em The Journal of Chemical Physics}, 97(6):4398--4412, 1992.

\bibitem{esposito2010efficiency}
Massimiliano Esposito, Ryoichi Kawai, Katja Lindenberg, and Christian Van~den
  Broeck.
\newblock Efficiency at maximum power of low-dissipation carnot engines.
\newblock {\em Physical review letters}, 105(15):150603, 2010.

\bibitem{quan2007quantum}
HT~Quan, Yu-xi Liu, CP~Sun, and Franco Nori.
\newblock Quantum thermodynamic cycles and quantum heat engines.
\newblock {\em Physical Review E}, 76(3):031105, 2007.

\bibitem{allahverdyan2008work}
Armen~E Allahverdyan, Ramandeep~S Johal, and Guenter Mahler.
\newblock Work extremum principle: Structure and function of quantum heat
  engines.
\newblock {\em Physical Review E}, 77(4):041118, 2008.

\bibitem{k299}
Raam Uzdin, Amikam Levy, and Ronnie Kosloff.
\newblock Quantum equivalence and quantum signatures in heat engines.
\newblock {\em Phys. Rev. X}, 5:031044, 2015.

\bibitem{chernoff1974product}
Paul~R Chernoff.
\newblock {\em Product formulas, nonlinear semigroups, and addition of
  unbounded operators}, volume 140.
\newblock American Mathematical Soc., 1974.

\bibitem{k152}
Tova Feldmann and Ronnie Kosloff.
\newblock Performance of discrete heat engines and heat pumps in finite time.
\newblock {\em Physical Review E}, 61(5):4774, 2000.

\bibitem{k243}
Yair Rezek, Peter Salamon, Karl~Heinz Hoffmann, and Ronnie Kosloff.
\newblock The quantum refrigerator: The quest for absolute zero.
\newblock {\em EPL (Europhysics Letters)}, 85(3):30008, 2009.

\bibitem{k24}
Ronnie Kosloff.
\newblock A quantum mechanical open system as a model of a heat engine.
\newblock {\em The Journal of Chemical Physics}, 80(4):1625--1631, 1984.

\bibitem{k169}
Jos{\'e}~P Palao, Ronnie Kosloff, and Jeffrey~M Gordon.
\newblock Quantum thermodynamic cooling cycle.
\newblock {\em Physical Review E}, 64(5):056130, 2001.

\bibitem{tonner2005autonomous}
Friedemann Tonner and G{\"u}nter Mahler.
\newblock Autonomous quantum thermodynamic machines.
\newblock {\em Physical Review E}, 72(6):066118, 2005.

\bibitem{linden2010small}
Noah Linden, Sandu Popescu, and Paul Skrzypczyk.
\newblock How small can thermal machines be? the smallest possible
  refrigerator.
\newblock {\em Physical review letters}, 105(13):130401, 2010.

\bibitem{k310}
Amikam Levy, Lajos Di\'osi, and Ronnie Kosloff.
\newblock Quantum flywheel.
\newblock {\em Phys. Rev. A}, 93:052119, 2016.

\bibitem{k272}
Amikam Levy and Ronnie Kosloff.
\newblock Quantum absorption refrigerator.
\newblock {\em Physical Reveiew Letters}, 108:070604, 2012.

\bibitem{saira2007heat}
Olli-Pentti Saira, Matthias Meschke, Francesco Giazotto, Alexander~M Savin,
  Mikko M{\"o}tt{\"o}nen, and Jukka~P Pekola.
\newblock Heat transistor: Demonstration of gate-controlled electronic
  refrigeration.
\newblock {\em Physical review letters}, 99(2):027203, 2007.

\bibitem{skrzypczyk2011smallest}
Paul Skrzypczyk, Nicolas Brunner, Noah Linden, and Sandu Popescu.
\newblock The smallest refrigerators can reach maximal efficiency.
\newblock {\em Journal of Physics A: Mathematical and Theoretical},
  44(49):492002, 2011.

\bibitem{palao13}
Luis~A Correa, Jos{\'e}~P Palao, Gerardo Adesso, and Daniel Alonso.
\newblock Performance bound for quantum absorption refrigerators.
\newblock {\em Physical Review E}, 87(4):042131, 2013.

\bibitem{k289}
Ronnie Kosloff and Amikam Levy.
\newblock Quantum heat engines and refrigerators: Continuous devices.
\newblock {\em Annual Review of Physical Chemistry}, 65:365--393, 2014.

\bibitem{maslennikov2017quantum}
Gleb Maslennikov, Shiqian Ding, Roland Hablutzel, Jaren Gan, Alexandre Roulet,
  Stefan Nimmrichter, Jibo Dai, Valerio Scarani, and Dzmitry Matsukevich.
\newblock Quantum absorption refrigerator with trapped ions.
\newblock {\em arXiv preprint arXiv:1702.08672}, 2017.

\bibitem{paz2013}
Esteban~A Martinez and Juan~Pablo Paz.
\newblock Dynamics and thermodynamics of linear quantum open systems.
\newblock {\em Physical review letters}, 110(13):130406, 2013.

\bibitem{clausius1850}
Rudolf Clausius.
\newblock {\"U}ber die bewegende kraft der w{\"a}rme und die gesetze, welche
  sich daraus f{\"u}r die w{\"a}rmelehre selbst ableiten lassen.
\newblock {\em Annalen der Physik}, 155(3):368--397, 1850.

\bibitem{correa2014quantum}
Luis~A Correa, Jos{\'e}~P Palao, Daniel Alonso, and Gerardo Adesso.
\newblock Quantum-enhanced absorption refrigerators.
\newblock {\em Scientific reports}, 4:srep03949, 2014.

\bibitem{mu2017qubit}
Anqi Mu, Bijay~Kumar Agarwalla, Gernot Schaller, and Dvira Segal.
\newblock Qubit absorption refrigerator at strong coupling.
\newblock {\em New Journal of Physics}, 19(12):123034, 2017.

\bibitem{nerst06}
{W. Nernst}.
\newblock {Ueber die Berechnung chemischer Gleichgewichte aus thermischen
  Messungen}.
\newblock {\em {Nachr. Kgl. Ges. Wiss. Go\"tt.}}, 1:40, 1906.

\bibitem{landsberg89}
{P. T. Landsberg}.
\newblock A comment on nernst's theorem.
\newblock {\em J. Phys A: Math.Gen.}, 22:139, 1989.

\bibitem{k278}
Amikam Levy, Robert Alicki, and Ronnie Kosloff.
\newblock Comment on ``cooling by heating: Refrigeration powered by photons''.
\newblock {\em Physical Review Letters}, 109:248901, 2012.

\bibitem{nerst18}
{W. Nernst}.
\newblock {\em {The theoretical and experimental bases of the New Heat Theorem
  Ger., Die theoretischen und experimentellen Grundlagen des neuen
  Wa\"rmesatzes}}.
\newblock W. Knapp, Halle, 1918.

\bibitem{masanes2017general}
Llu{\'\i}s Masanes and Jonathan Oppenheim.
\newblock A general derivation and quantification of the third law of
  thermodynamics.
\newblock {\em Nature Communications}, 8, 2017.

\bibitem{freitas2017fundamental}
Nahuel Freitas and Juan~Pablo Paz.
\newblock Fundamental limits for cooling of linear quantum refrigerators.
\newblock {\em Physical Review E}, 95(1):012146, 2017.

\bibitem{gelbwaser2013minimal}
David Gelbwaser-Klimovsky, Robert Alicki, and Gershon Kurizki.
\newblock Minimal universal quantum heat machine.
\newblock {\em Physical Review E}, 87(1):012140, 2013.

\bibitem{zhou2010minimal}
Yun Zhou and Dvira Segal.
\newblock Minimal model of a heat engine: Information theory approach.
\newblock {\em Physical Review E}, 82(1):011120, 2010.

\bibitem{gelbwaser2013work}
David Gelbwaser-Klimovsky, Robert Alicki, and Gershon Kurizki.
\newblock Work and energy gain of heat-pumped quantized amplifiers.
\newblock {\em EPL (Europhysics Letters)}, 103(6):60005, 2013.

\bibitem{boukobza2007three}
E~Boukobza and DJ~Tannor.
\newblock Three-level systems as amplifiers and attenuators: A thermodynamic
  analysis.
\newblock {\em Physical review letters}, 98(24):240601, 2007.

\bibitem{li2017quantum}
Sheng-Wen Li, Moochan~B Kim, Girish~S Agarwal, and Marlan~O Scully.
\newblock Quantum statistics of a single-atom scovil--schulz-dubois heat
  engine.
\newblock {\em Physical Review A}, 96(6):063806, 2017.

\bibitem{benenti2016fundamental}
Giuliano Benenti, Giulio Casati, Keiji Saito, and Robert~S Whitney.
\newblock Fundamental aspects of steady-state conversion of heat to work at the
  nanoscale.
\newblock {\em arXiv preprint arXiv:1608.05595}, 2016.

\bibitem{alicki2017thermodynamic}
Robert Alicki, David Gelbwaser-Klimovsky, and Alejandro Jenkins.
\newblock A thermodynamic cycle for the solar cell.
\newblock {\em Annals of Physics}, 378:71--87, 2017.

\bibitem{alicki2015solar}
Robert Alicki, David Gelbwaser-Klimovsky, and Krzysztof Szczygielski.
\newblock Solar cell as a self-oscillating heat engine.
\newblock {\em Journal of Physics A: Mathematical and Theoretical},
  49(1):015002, 2015.

\bibitem{alicki2016thermoelectric}
Robert Alicki.
\newblock Thermoelectric generators as self-oscillating heat engines.
\newblock {\em Journal of Physics A: Mathematical and Theoretical},
  49(8):085001, 2016.

\bibitem{ee2017organic}
Harrison Lee, Jiaying Wu, Jeremy Barbe, Sagar~Motilal Jain, Sebastian Wood,
  Emily Speller, Zhe Li, Fernando~Araujo de~Castro, James Durrant, and Wing
  Tsoi.
\newblock Organic photovoltaic cell--a promising indoor light harvester for
  self-sustainable electronics.
\newblock {\em Journal of Materials Chemistry A}, 2017.

\bibitem{bakulin2015mode}
Artem~A Bakulin, Robert Lovrincic, Xi~Yu, Oleg Selig, Huib~J Bakker, Yves~LA
  Rezus, Pabitra~K Nayak, Alexandr Fonari, Veaceslav Coropceanu, Jean-Luc
  Br{\'e}das, et~al.
\newblock Mode-selective vibrational modulation of charge transport in organic
  electronic devices.
\newblock {\em Nature communications}, 6, 2015.

\bibitem{chin2013role}
AW~Chin, J~Prior, R~Rosenbach, F~Caycedo-Soler, SF~Huelga, and MB~Plenio.
\newblock The role of non-equilibrium vibrational structures in electronic
  coherence and recoherence in pigment-protein complexes.
\newblock {\em Nature Physics}, 9(2):113--118, 2013.

\bibitem{nielsen2002quantum}
Michael~A Nielsen and Isaac Chuang.
\newblock Quantum computation and quantum information, 2002.

\bibitem{Landauer:1961}
Rolf Landauer.
\newblock Irreversibility and heat generation in the computing process.
\newblock {\em IBM journal of research and development}, 5(3):183--191, 1961.

\bibitem{Bennett:2003}
Charles~H Bennett.
\newblock Notes on landauer's principle, reversible computation, and maxwell's
  demon.
\newblock {\em Studies In History and Philosophy of Science Part B: Studies In
  History and Philosophy of Modern Physics}, 34(3):501--510, 2003.

\bibitem{szilard1925}
Leo Szilard.
\newblock {\"U}ber die ausdehnung der ph{\"a}nomenologischen thermodynamik auf
  die schwankungserscheinungen.
\newblock {\em Zeitschrift f{\"u}r Physik A Hadrons and Nuclei},
  32(1):753--788, 1925.

\bibitem{szilard1929}
Leo Szilard.
\newblock {\"U}ber die entropieverminderung in einem thermodynamischen system
  bei eingriffen intelligenter wesen.
\newblock {\em Zeitschrift f{\"u}r Physik A Hadrons and Nuclei},
  53(11):840--856, 1929.

\bibitem{Sagawa:2012}
Takahiro Sagawa.
\newblock Thermodynamics of information processing in small systems*.
\newblock {\em Progress of theoretical physics}, 127(1):1--56, 2012.

\bibitem{Toyabe}
Shoichi Toyabe, Takahiro Sagawa, Masahito Ueda, Eiro Muneyuki, and Masaki Sano.
\newblock Experimental demonstration of information-to-energy conversion and
  validation of the generalized jarzynski equality.
\newblock {\em Nature Physics}, 6(12):988--992, 2010.

\bibitem{Lutz}
Antoine B{\'e}rut, Artak Arakelyan, Artyom Petrosyan, Sergio Ciliberto, Raoul
  Dillenschneider, and Eric Lutz.
\newblock Experimental verification of landauer/'s principle linking
  information and thermodynamics.
\newblock {\em Nature}, 483(7388):187--189, 2012.

\bibitem{pekola2015towards}
Jukka~P Pekola.
\newblock Towards quantum thermodynamics in electronic circuits.
\newblock {\em Nature Physics}, 11(2):118--123, 2015.

\bibitem{Norton}
John~D Norton.
\newblock Eaters of the lotus: Landauer's principle and the return of maxwell's
  demon.
\newblock {\em Studies In History and Philosophy of Science Part B: Studies In
  History and Philosophy of Modern Physics}, 36(2):375--411, 2005.

\bibitem{Shenker}
Orly~R Shenker.
\newblock Logic and entropy.
\newblock 2000.

\bibitem{alicki2014information}
Robert Alicki.
\newblock Information is not physical.
\newblock {\em arXiv preprint arXiv:1402.2414}, 2014.

\bibitem{Feyerabend}
Paul Feyerabend.
\newblock Consolations for the specialist.
\newblock {\em Criticism and the Growth of Knowledge}, 4:197--229, 1970.

\bibitem{horodecki2013fundamental}
Micha{\l} Horodecki and Jonathan Oppenheim.
\newblock Fundamental limitations for quantum and nanoscale thermodynamics.
\newblock {\em Nature communications}, 4:2059, 2013.

\bibitem{gour2015resource}
Gilad Gour, Markus~P M{\"u}ller, Varun Narasimhachar, Robert~W Spekkens, and
  Nicole~Yunger Halpern.
\newblock The resource theory of informational nonequilibrium in
  thermodynamics.
\newblock {\em Physics Reports}, 583:1--58, 2015.

\bibitem{goold2016role}
John Goold, Marcus Huber, Arnau Riera, L{\'\i}dia del Rio, and Paul Skrzypczyk.
\newblock The role of quantum information in thermodynamics?a topical review.
\newblock {\em Journal of Physics A: Mathematical and Theoretical},
  49(14):143001, 2016.

\bibitem{vinjanampathy2016quantum}
Sai Vinjanampathy and Janet Anders.
\newblock Quantum thermodynamics.
\newblock {\em Contemporary Physics}, 57(4):545--579, 2016.

\bibitem{korzekwa2016extraction}
Kamil Korzekwa, Matteo Lostaglio, Jonathan Oppenheim, and David Jennings.
\newblock The extraction of work from quantum coherence.
\newblock {\em New Journal of Physics}, 18(2):023045, 2016.

\bibitem{goold16}
John Goold, Marcus Huber, Arnau Riera, Lídia del Rio, and Paul Skrzypczyk.
\newblock The role of quantum information in thermodynamics: a topical review.
\newblock {\em Journal of Physics A: Mathematical and Theoretical},
  49(14):143001, 2016.

\bibitem{brandao2015second}
Fernando Brandao, Micha{\l} Horodecki, Nelly Ng, Jonathan Oppenheim, and
  Stephanie Wehner.
\newblock The second laws of quantum thermodynamics.
\newblock {\em Proceedings of the National Academy of Sciences},
  112(11):3275--3279, 2015.

\bibitem{renyi1961measures}
Alfr{\'e}d R{\'e}nyi et~al.
\newblock On measures of entropy and information.
\newblock In {\em Proceedings of the Fourth Berkeley Symposium on Mathematical
  Statistics and Probability, Volume 1: Contributions to the Theory of
  Statistics}. The Regents of the University of California, 1961.

\bibitem{talkner2007fluctuation}
Peter Talkner, Eric Lutz, and Peter H{\"a}nggi.
\newblock Fluctuation theorems: Work is not an observable.
\newblock {\em Physical Review E}, 75(5):050102, 2007.

\bibitem{roncaglia2014work}
Augusto~J Roncaglia, Federico Cerisola, and Juan~Pablo Paz.
\newblock Work measurement as a generalized quantum measurement.
\newblock {\em Physical review letters}, 113(25):250601, 2014.

\bibitem{sampaio2017impossible}
Rui Sampaio, Samu Suomela, Tapio Ala-Nissila, Janet Anders, and Thomas Philbin.
\newblock The impossible quantum work distribution.
\newblock {\em arXiv preprint arXiv:1707.06159}, 2017.

\bibitem{gelbwaser2017thermodynamic}
David Gelbwaser-Klimovsky and Al{\'a}n Aspuru-Guzik.
\newblock On thermodynamic inconsistencies in several photosynthetic and solar
  cell models and how to fix them.
\newblock {\em Chemical Science}, 8(2):1008--1014, 2017.

\bibitem{boukobza2013breaking}
E~Boukobza and H~Ritsch.
\newblock Breaking the carnot limit without violating the second law: A
  thermodynamic analysis of off-resonant quantum light generation.
\newblock {\em Physical Review A}, 87(6):063845, 2013.

\bibitem{alicki2017gkls}
Robert Alicki.
\newblock From the gkls equation to the theory of solar and fuel cells.
\newblock {\em Open Systems \& Information Dynamics}, 24(03):1740007, 2017.

\bibitem{k11}
Ronnie Kosloff and Stuart~A Rice.
\newblock The influence of quantization on the onset of chaos in hamiltonian
  systems: the kolmogorov entropy interpretation.
\newblock {\em The Journal of Chemical Physics}, 74(2):1340--1349, 1981.

\bibitem{deutsch1991quantum}
Josh~M Deutsch.
\newblock Quantum statistical mechanics in a closed system.
\newblock {\em Physical Review A}, 43(4):2046, 1991.

\bibitem{srednicki1994chaos}
Mark Srednicki.
\newblock Chaos and quantum thermalization.
\newblock {\em Physical Review E}, 50(2):888, 1994.

\bibitem{dyson1962statistical}
Freeman~J Dyson.
\newblock Statistical theory of the energy levels of complex systems. i.
\newblock {\em Journal of Mathematical Physics}, 3(1):140--156, 1962.

\bibitem{kim2014testing}
Hyungwon Kim, Tatsuhiko~N Ikeda, and David~A Huse.
\newblock Testing whether all eigenstates obey the eigenstate thermalization
  hypothesis.
\newblock {\em Physical Review E}, 90(5):052105, 2014.

\bibitem{steinigeweg2014pushing}
Robin Steinigeweg, Abdelah Khodja, Hendrik Niemeyer, Christian Gogolin, and
  Jochen Gemmer.
\newblock Pushing the limits of the eigenstate thermalization hypothesis
  towards mesoscopic quantum systems.
\newblock {\em Physical review letters}, 112(13):130403, 2014.

\bibitem{ikeda2013finite}
Tatsuhiko~N Ikeda, Yu~Watanabe, and Masahito Ueda.
\newblock Finite-size scaling analysis of the eigenstate thermalization
  hypothesis in a one-dimensional interacting bose gas.
\newblock {\em Physical Review E}, 87(1):012125, 2013.

\bibitem{alba2015eigenstate}
Vincenzo Alba.
\newblock Eigenstate thermalization hypothesis and integrability in quantum
  spin chains.
\newblock {\em Physical Review B}, 91(15):155123, 2015.

\bibitem{zhao2016dynamics}
P~Zhao, H~De~Raedt, S~Miyashita, F~Jin, and K~Michielsen.
\newblock Dynamics of open quantum spin systems: An assessment of the quantum
  master equation approach.
\newblock {\em Physical Review E}, 94(2):022126, 2016.

\bibitem{de2017relaxation}
Hans De~Raedt, Fengping Jin, Mikhail~I Katsnelson, and Kristel Michielsen.
\newblock Relaxation, thermalization, and markovian dynamics of two spins
  coupled to a spin bath.
\newblock {\em Physical Review E}, 96(5):053306, 2017.

\end{thebibliography}

\bibliographystyle{unsrt}

\end{document}